%% file: main.tex
\documentclass[
amsmath,
amssymb,
groupedaddress,
notitlepage,
floatfix,
aps,
prf,
dvipsnames,
]{revtex4-2}
\usepackage{CJK}
\usepackage[letterpaper,margin=1in]{geometry}
\usepackage{graphicx}
\usepackage{dcolumn}
\usepackage{bm}
\usepackage[FIGTOPCAP]{subfigure}
\usepackage{CJK}
\usepackage{xcolor}
\usepackage{rotating}
\usepackage{braket}
\usepackage{float}
\usepackage{hyperref}
\usepackage{protosem}

\definecolor{rbctop}{RGB}{0,103,154}
\definecolor{rbcbottom}{RGB}{152,0,76}
\newcommand{\goldfish}{%
\begin{minipage}[c][1em]{3.ex}{\vspace*{0.1em}
\begin{tabular}{@{}c@{}} {\color{rbctop} \textproto{\Amem}}\\[-1.6ex] \textproto{\Adaleth}\\[-1.6ex] {\color{rbcbottom} \textproto{\Amem}}\end{tabular}
}\end{minipage}}
\newcommand{\vect}{\boldsymbol}
\renewcommand{\vec}{\boldsymbol}
\begin{document}   

\bibliographystyle{apsrev4-1}

\begin{CJK*}{UTF8}{gbsn}
\title{The Transition from Wall Modes to Multimodality in Liquid Gallium Magnetoconvection}
\author{Yufan Xu (徐宇凡)} 
\email{yufanxu@g.ucla.edu}
\affiliation{Department of Earth, Planetary, and Space Science, University of California, Los Angeles, California 90095, USA}
\author{Susanne Horn (苏珊娜·霍恩)}
\affiliation{Centre for Fluid and Complex Systems, Coventry University, CV1 2NL Coventry, UK}
\author{Jonathan M. Aurnou（乔纳森·奥诺）}
\affiliation{Department of Earth, Planetary, and Space Science, University of California, Los Angeles, California 90095, USA } 
\date{\today}
\maketitle
\end{CJK*}
\begin{abstract}
Coupled laboratory-numerical experiments of Rayleigh-B\'enard convection (RBC) in liquid gallium subject to a vertical magnetic field are presented. The experiments are carried out in two cylindrical containers with diameter-to-height aspect ratio $\Gamma = 1.0$ and $2.0$ at varying thermal forcing (Rayleigh numbers $10^5 \lesssim Ra \lesssim 10^8$) and magnetic field strength (Chandrasekhar numbers $0\lesssim Ch \lesssim 3\times 10^5$). Laboratory measurements and numerical simulations confirm that magnetoconvection in our finite cylindrical tanks onsets via non-drifting wall-attached modes, in good agreement with asymptotic predictions for a semi-infinite domain. With increasing supercriticality, the experimental and numerical thermal measurements and the numerical velocity data reveal transitions between wall mode states with different azimuthal mode numbers and between wall-dominated convection to wall and interior multimodality. These transitions are also reflected in the heat transfer data, which combined with previous studies, connect onset to supercritical turbulent behaviors in liquid metal magnetoconvection over a large parameter space. The gross heat transfer behaviors between magnetoconvection and rotating convection in liquid metals are compared and discussed.
\end{abstract}
\maketitle

\section{Introduction} \label{sec:1_intro}
\input{sections/1_intro}

\section{Control parameters and linear prediction} \label{sec:2_theory}
\input{sections/2_theory}

\section{Methods} \label{sec:3_method}
\input{sections/3_method}

\section{Results} \label{sec:4_results}
\input{sections/4_results}

\section{Discussion} \label{sec:5_discussion}
\input{sections/5_discussion}

\section*{Acknowledgments}

We gratefully acknowledge the support of the NSF Geophysics Program (EAR awards \#1853196 and \#2143939) and S.H. also thanks the EPSRC (grant \#EP/V047388/1). Thanks also go to Jewel Abbate and Taylor Lonner for providing assistance with the experimental setup of RoMag, and to Meredith Plumley for generously sharing her extracted $Nu$-$Ra$ data from King \& Aurnou (2015) and Burr \& Muller (2011). 


\bibliography{bib}

\clearpage
\section{Appendix} \label{sec:appendix}
\input{sections/appendix}

\end{document}

%% file: sections/1_intro.tex
%
%
Convection influenced by ambient magnetic fields is called magnetoconvection (MC), which arises in many areas of fluid dynamics. In geophysics and planetary physics, motions of turbulent convective flows in planetary liquid metal outer cores generate planetary-scale magnetic fields via dynamo processes. Studying the effects of magnetic fields in MC is essential to understand these processes \citep[e.g.,][]{jones2011planetary,roberts2013genesis, aurnou2017cross,moffatt2019self}. In astrophysics, MC is associated with the sunspot umbra on the outer layer of the Sun and other stars \citep[e.g.,][]{proctor1982magnetoconvection,schussler2006magnetoconvection,rempel2009radiative}. Furthermore, MC has an essential role in numerous industrial and engineering applications including but not limited to liquid metal batteries \citep{kelley2018fluid, cheng2022laboratory}, crystal growth \citep{moreau1999fundamentals, rudolph2008travelling}, nuclear fusion liquid-metal cooling blanket designs \citep{barleon1991mhd,abdou2001exploration,salavy2007overview}, and induction heating and casting \citep{taberlet1985turbulent,davidson1999magnetohydrodynamics}. These systems have drastically different ratios between electromagnetic and inertial forces. Therefore, it is crucial to investigate the effects of a wide range of magnetic forces in MC systems. 

The canonical model of MC is a convection system with an electrically-conducting fluid layer heated from below, cooled from above, and in the presence of an external vertical magnetic field \cite{chandrasekhar1961hydrodynamic,yan2019heat}. It is most fundamentally understood in an extended plane layer geometry \citep[e.g.,][]{chandrasekhar1961hydrodynamic,nakagawa1955experiment,yan2019heat}. But there is an increasing interest in MC systems with defined sidewall boundaries because of their many experimental and industrial applications. Various numerical and laboratory studies have been carried out in rectangular \citep{schussler2006magnetoconvection, liu2018wall} and cylindrical geometries \citep{cioni2000effect,akhmedagaev2020turbulent,zurner2019combined,zurner2020flow}. 

In weakly supercritical, near-onset regimes, MC systems tend to develop steady wall modes in the sidewall Shercliff boundary layer \citep{houchens2002rayleigh,busse2008asymptotic,liu2018wall,zurner2020flow,akhmedagaev2020turbulent} while the bulk remains quiescent. As the magnetoconvective supercriticality increases, convective flows self-organise into multi-cellular bulk flow structures \citep{yan2019heat, zurner2020flow}. Eventually, at very large supercriticalities, the buoyancy forces dominate and magnetic field effects become subdominant. Large-scale circulations (LSCs) then form and the heat and momentum transfer asymptote to that of turbulent RBC \citep[e.g.][]{lim2019quasistatic,zurner2020flow,xu2022thermoelectric,grannan2022experimental}. 

The pathway from the onset of convection to fully developed turbulence in liquid metal MC is not well characterised. To address this deficit, we present a suite of laboratory-numerical coupled MC experiments in liquid gallium to investigate how MC transitions from near-onset wall modes to turbulent multimodality in cylindrical cells. The paper is organised as follows. Section \ref{sec:2_theory} introduces control parameters, and reviews established onset predictions for the magnetoconvection system. Section \ref{sec:3_method} presents our experimental setup, numerical schemes, diagnostics, and the physical properties of liquid gallium. Section \ref{sec:4_results} compares different theoretical onsets and observations of the transition to multimodality in a survey with fixed magnetic field strength and varying convective vigor. Section \ref{sec:5_discussion} shows heat transfer results combining  previous studies of MC and compares the gross heat transfer behaviors between liquid metal magnetoconvection and liquid metal rotating convection systems.

%% file: sections/2_theory.tex
%
%
Laboratory-scale liquid metal magnetoconvection usually has a negligible induced magnetic field $\boldsymbol{b}$ with respect to the external applied magnetic field $\vec{B}_0$, so that $|\vec{b}|\ll |\vec{B}_0|$. Moreover, any induced field is considered temporally invariant, $\partial_t \boldsymbol{b}\approx 0$. The magnetic Reynolds number, defined as $Rm = UH/\eta$ \citep{julien1996rapidly,glazier1999evidence,davidson_2016,akhmedagaev2020turbulent}, is well below unity, where $U$ and $H$ are the characteristic velocity and length scales, respectively, and $\eta$ is the magnetic diffusivity. This parameter represents the ratio between induction and diffusion of the magnetic field. Thus, the so-called low-$Rm$ quasistatic approximation is valid in most liquid metal experimental and industrial applications \citep{sarris2006limits,knaepen2008magnetohydrodynamic,davidson_2016,cioni2000effect}. In the quasistatic limit, $Rm$ and magnetic Prandtl number $Pm$ formally drop out of the problem, so it is not necessary to solve the magnetic induction equation explicitly, and the system is greatly simplified. In addition, the Oberbeck-Boussinesq approximation is commonly applied in the governing equations for liquid metal MC systems \citep[e.g.][]{cioni2000effect,liu2018wall,vogt2018jump,yan2019heat,xu2022thermoelectric}.

Four nondimensional control parameters govern quasistatic Oberbeck-Boussinesq magnetoconvection \citep{liu2018wall,yan2019heat,xu2022thermoelectric}. The Prandtl number $Pr$ describes the thermo-mechanical properties of the fluid, 
\begin{equation}
    Pr =\frac{\nu}{\kappa}, 
\end{equation}
where $\nu$ is the kinematic viscosity and $\kappa$ is the thermal diffusivity. In this study, $Pr \approx 0.027$ for liquid gallium. The Rayleigh number $Ra$ characterises the buoyancy forcing with respect to thermo-viscous diffusion and is defined as
\begin{equation}
   Ra =  \frac{\alpha g \Delta T H^3}{\kappa \nu}.
\end{equation}
Here, $\alpha$ is the thermal expansion coefficient, $g$ is the magnitude of the vertically oriented ($\vec{\hat e_z}$) gravitational acceleration, $\Delta T$ is the bottom-to-top vertical temperature difference across the fluid layer, and the characteristic length scale is the layer height $H$. The Chandrasekhar number $Ch$ denotes the ratio of quasistatic Lorentz forces and viscous forces,
\begin{equation}
   Ch = \frac{\sigma B_0^2 H^2}{\rho_0 \nu},
\end{equation}
where $\sigma$ is the electric conductivity of the fluid, $B_0$ is the magnitude of the applied vertical magnetic field, and $\rho_0$ is the mean density of the fluid. The Chandrasekhar number is the square of the Hartmann number, $Ch = Ha^2$ \citep[e.g.][]{davidson_2016,moreau1999fundamentals,roberts1967introduction}. Additionally, the cylindrical container has a diameter-to-height aspect ratio
\begin{equation}
    \Gamma = \frac{D}{H},
\end{equation}
where $D$ is the diameter of the container. Here, $\Gamma$ is fixed to $1.0$ and $2.0$, respectively.

\begin{figure}[t]
  \centering
    \makebox[\textwidth][c] {\includegraphics[width=\textwidth]{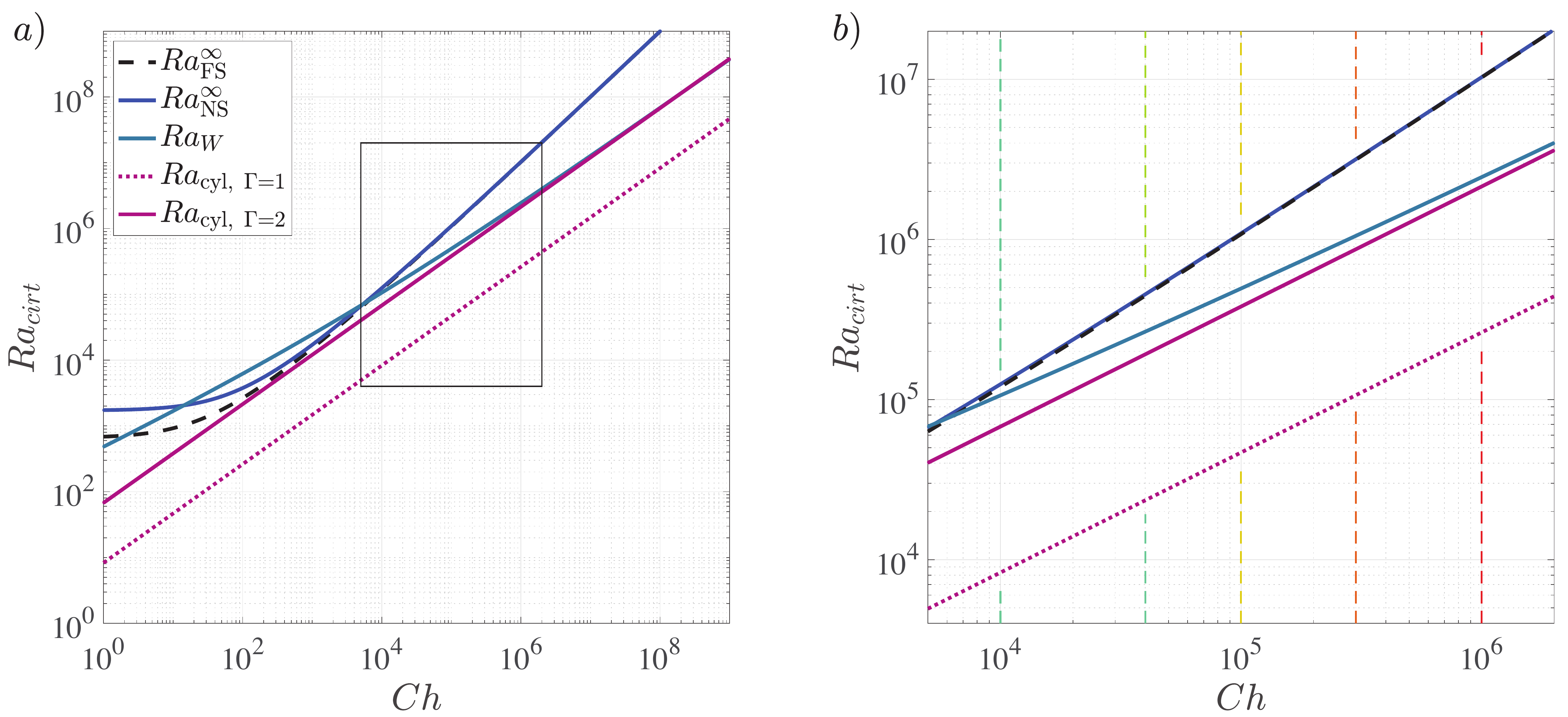}}
 	  \caption{a) Different prediction of critical Rayleigh number ($Ra_{crit}$) as a function of the Chandrasekhar number. The black dashed curve shows the $Ra_{cirt}$. For infinite plane MC system with free-slip boundaries at the top and bottom, $Ra_{crit} = Ra_{FS}^{\infty}$ \citep{chandrasekhar1961hydrodynamic}, as shown in the black dashed curve; for infinite plane MC system with no-slip boundaries, $Ra_{crit} = Ra_{NS}^{\infty}$ \citep{chandrasekhar1952xlvi}, as shown in the blue curve. Note that both $Ra_{FS}^{\infty}$ and $Ra_{NS}^{\infty}$ asymptote to $\pi^2Ch$ as $Ch \rightarrow \infty$; the grayish-blue curve shows that $Ra_{W}$ is the asymptotic $Ra_{crit}$ for the wall-mode onset in a half-infinite plane with a vertical boundary \cite{busse2008asymptotic}; the purple curve and the purple dotted curve are \citet{houchens2002rayleigh}'s predicted $Ra_{crit}$, namely $Ra_{cyl,\ \Gamma = 2}$ and $Ra_{cyl,\ \Gamma = 1}$, for MC in cylindrical containers with aspect ratio $1$ and $2$, respectively. b) The zoom-in view of the region circumscribed by the black rectangular box in panel a). The colored vertical dashed lines correspond to five $Ch$ numbers employed in our study. }
   \label{fig:onset}
\end{figure}

The onset of convection is controlled by the critical Rayleigh denoted as $Ra_{crit}$, which characterises the buoyancy forcing needed for a particular convective mode in the system \citep{plumley2019scaling}. Figure \ref{fig:onset} shows different $Ra_{crit}$ predictions. Linear analysis has shown that the convection driven by buoyancy forces must balance the viscous and Joule dissipation \citep{chandrasekhar1961hydrodynamic}. Thus, in general, the magnetic field inhibits the onset of the convection. \citet{chandrasekhar1961hydrodynamic} derived the onset for the bulk stationary magnetoconvection in an infinite plane layer ($\infty$). With free-slip (FS) boundaries on both ends, the dispersion relation expresses the marginal Rayleigh number $Ra_M$ as,
\begin{equation}
   Ra_M =\frac{\pi^2+a^2}{a^2} \left[ \left( \pi^2 + a^2 \right)^2 + \pi^2 Ch\right],
   \label{eq:FS}
\end{equation}
where $a$ is the characteristic cell aspect ratio \citep{davidson_2016}, defined as $a \equiv \pi H /L$, where $H$ is the height of the fluid layer, and $2L$ is the horizontal wavelength of the convection flow, assuming the form of two-dimensional rolls with each roll having diameter $L$. By minimizing equation (\ref{eq:FS}), setting $\partial Ra/\partial a = 0$, we obtain the critical Rayleigh number for the bulk stationary magnetoconvection in an infinite layer with free-slip boundaries on both ends, $Ra^{\infty}_{FS}$, and its critical mode number $a_{FS}$. In the limit of $ Ch \rightarrow \infty$, we have $Ra^{\infty}_{FS} \rightarrow \pi^2 Ch$, and $a_{FS} \rightarrow (\pi^4 Ch/2)^{1/6}$ \citep{chandrasekhar1961hydrodynamic,davidson_2016}.

Magnetoconvection with no-slip (NS) rigid horizontal boundaries has a dispersion relation \citep{chandrasekhar1952xlvi}
\begin{equation}
   Ra_M =\frac{\left(\pi^{2}+a^{2}\right)\left[\left(\pi^{2}+a^{2}\right)^{2}+\pi^{2} Ch\right]}{a^{2}\left[1-4 \pi^{2} \delta\left(q_{1}^{2}-q_{2}^{2}\right) /\left(\pi^{2}+q_{1}^{2}\right)\left(\pi^{2}+q_{2}^{2}\right)\right]}
   \label{eq:ns}
\end{equation}
where 
\begin{equation}
    q_1=\frac{1}{2}\left( \sqrt{Ch+4 a^{2}}+\sqrt{Ch}\right), \quad
    q_2=\frac{1}{2}\left( \sqrt{Ch+4 a^{2}}-\sqrt{Ch}\right),
\end{equation}
and
\begin{equation}
    \delta = \left(q_1 \tanh (q_1/2) - q_2 \tanh(q_2/2)\right)^{-1}.
\end{equation}
By assuming a single structure in the vertical direction and minimizing $Ra$ in (\ref{eq:ns}), we obtain the first approximation of critical $Ra$ of magnetoconvection with two rigid boundaries \cite{chandrasekhar1952xlvi}. Equation (\ref{eq:ns}) also predicts that $Ra^{\infty}_{NS}\rightarrow \pi^2 Ch$ and $a_{NS} \rightarrow (\pi^4 Ch/2)^{1/6}$ with $Ch\rightarrow \infty$. Thus, both critical Rayleigh numbers with free-slip boundaries $Ra^{\infty}_{FS}$ and no-slip boundaries $Ra^{\infty}_{NS}$ asymptote to $\pi^2 Ch$ above $Ch \gtrsim 10^4$, as shown in figure \ref{fig:onset}a). These asymptotic bulk onset predictions agree with previous experimental results \citep{nakagawa1955experiment,cioni2000effect,yan2019heat,zurner2020flow}. 

\citet{busse2008asymptotic} theoretically analysed the side wall modes in MC and derived an asymptotic solution along a straight vertical sidewall in a semi-infinite domain with free-slip top-bottom boundaries. The critical Rayleigh number $Ra_{W}$ for these so-called magnetowall modes is,
\begin{equation}
   Ra_{W} = 3\pi^2\sqrt{3\pi/2}\left(1+3Ch^{-1/4}\sqrt{3\pi/2}\right) Ch^{3/4}.
   \label{eq:Busse}
\end{equation}
The asymptotic onset of the wall modes is generally lower than the onset in the bulk fluid at large $Ch$, since $Ch^{3/4}\ll Ch$ as $Ch \rightarrow \infty$. These magnetowall modes are non-drifting and extend into the fluid bulk with a distance that scales as the magnetic boundary layer thickness, which scales with the Shercliff boundary layer thickness $\delta_{Sh} \sim Ch^{-1/4}$ \citep{shercliff1953steady, liu2018wall}. The stationary wall modes of MC differ from those found in rotating convection, where wall modes drift in azimuth \citep{ecke1992hopf, herrmann1993asymptotic, horn2017prograde}.

\citet{houchens2002rayleigh} performed a hybrid linear stability analysis combining the analytical solution for the $\delta_{Ha} \sim Ch^{-1/2}$ Hartmann layers \citep{hartmann1937hg} at top-bottom boundaries and numerical solutions for the rest of the domain in $\Gamma = 1$ and $2$ cylindrical geometries. They also presented a linear asymptotic analysis for large $Ch$. Their asymptotic solutions for critical $Ra$ for $\Gamma = 1$ and $2$ are, respectively,
\begin{equation}
   Ra_{cyl,\Gamma=1} = 8.302 Ch^{3/4};\quad Ra_{cyl,\Gamma=2} = 67.748 Ch^{3/4}.
   \label{eq:cyl}
\end{equation}
Figure \ref{fig:onset} summarises all the critical Rayleigh predictions mentioned above. \citet{houchens2002rayleigh}'s $Ra_{cyl,\Gamma = 1}$ values (marked by the purple dashed line) are approximately an order of magnitude lower than the rest of the onset predictions that are not aspect-ratio dependent. To test the validity of these predictions, we combine laboratory experiments and direct numerical simulations (DNS) to investigate the five different $Ch$ shown by the vertical dashed lines in figure \ref{fig:onset}b). The values of these five $Ch$ numbers and their corresponding critical $Ra$ are summarised in table \ref{table0}. 
\begin{table}[ht]
\begin{ruledtabular}
\begin{tabular}{l|ccccc}
$\bm{Ch}$    & $\bm{Ra_{\mathrm{FS}}^\infty} $  & $\bm{Ra_{\mathrm{NS}}^\infty}$ & $\bm{Ra_{W}} $ & $\bm{Ra_{\mathrm{cyl},\ \Gamma =1}}$ & $\bm{Ra_{\mathrm{cyl},\ \Gamma =2}}$\\[0.5ex]
\hline\\[-1.5ex]
$1\times 10^4$   & $ 1.20\times 10^5$    &$ 1.25\times 10^5$      & $ 1.06\times 10^5$   & $8.30\times 10^3 $ & $6.77\times 10^4$ \\[1.5ex]
$4\times 10^4$      & $ 4.46\times 10^5$   & $ 4.54\times 10^5$      & $2.66\times 10^5 $    & $ 2.35\times 10^4$  & $1.92\times 10^5 $ \\[1.5ex]
$1\times 10^5$      & $ 1.08\times 10^6$   & $ 1.09\times 10^6$   &    $4.94\times 10^5 $   & $ 4.67\times 10^4$ & $	3.81\times 10^5 $  \\[1.5ex]
$3\times 10^5$      & $ 3.15\times 10^6$   & $ 3.17\times 10^6$  &    $	1.05\times 10^6$   & $	1.06\times 10^5 $ & $8.68\times 10^5 $ \\[1.5ex]
$1\times 10^6$    & $1.03\times 10^7 $   & $ 1.03\times 10^7$   &    $2.45\times 10^6 $   & $2.63\times 10^5 $ & $	2.14\times 10^6 $ \\[1.5ex]
\end{tabular}
\end{ruledtabular}
\caption{Values of different predicted critical $Ra$ at $Ch = \{10^4$, $4\times 10^4$, $10^5$, $3\times 10^5$, $10^6\}$, which have been examined experimentally.}
\label{table0}
\end{table}
%

%% file: sections/3_method.tex
\begin{figure}[t]
   \centering
    \makebox[\textwidth][c] {\includegraphics[width=0.95\columnwidth]{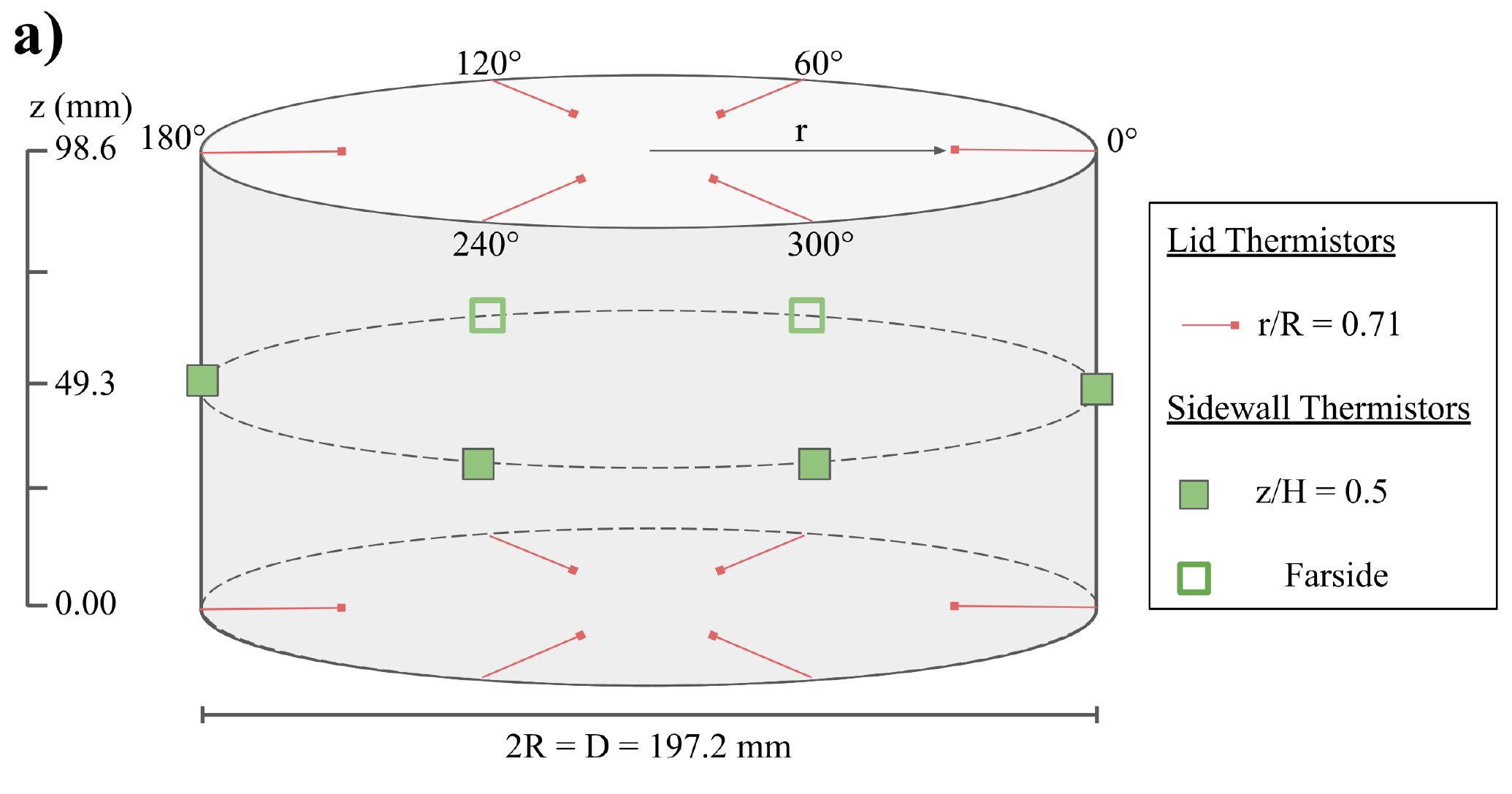}}
    \makebox[\textwidth][c]{\includegraphics[width=0.95\columnwidth]{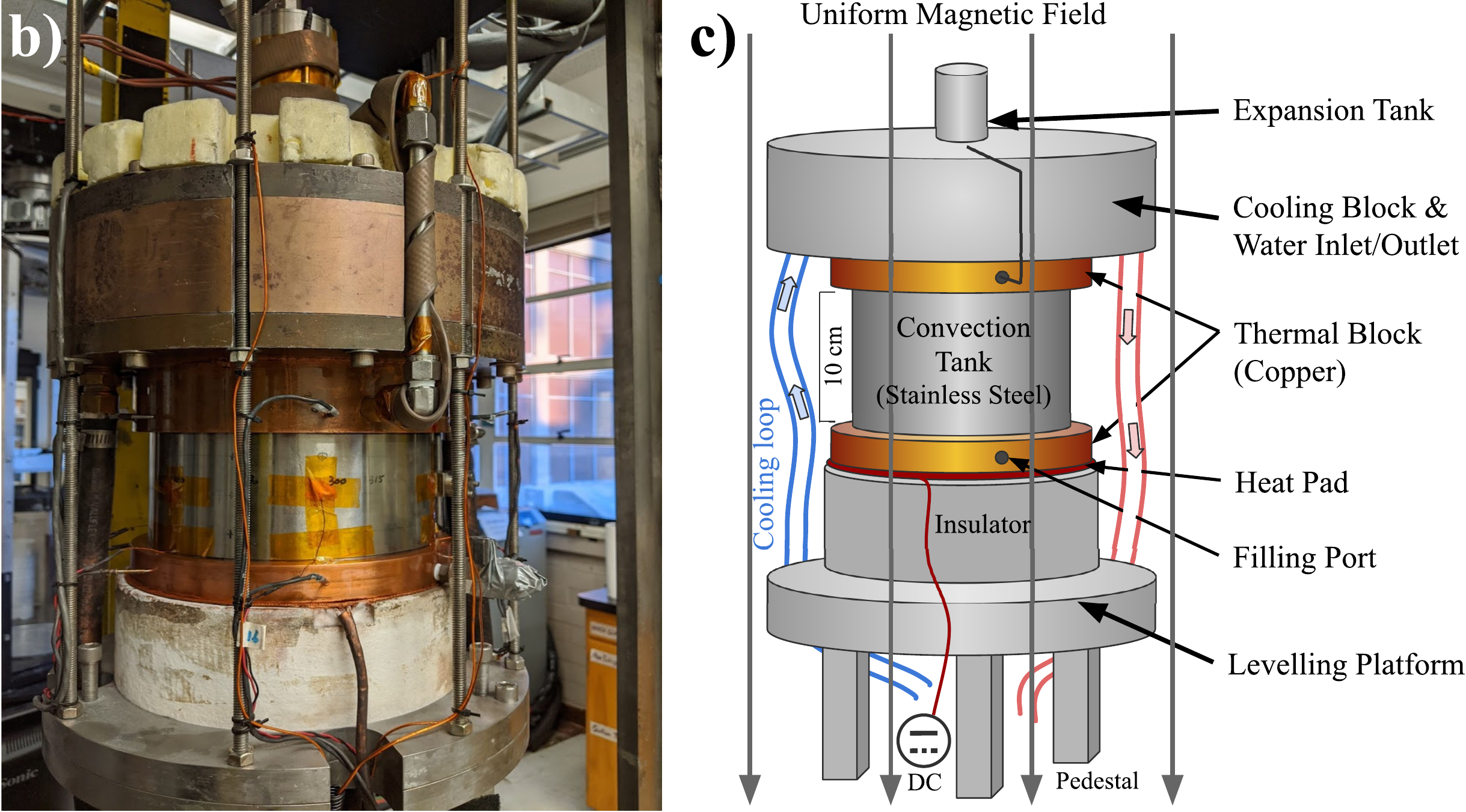}}
 	  \caption{ a) Thermometry schematic for the $\Gamma = 2R/H = 2.0$ tank ($R = 98.6\, \mathrm{mm}$): there are six thermistors in the top lid at $0.71R$, six thermistors at the midplane sidewall, and six thermistors in the bottom lid at $0.71R$. These thermistors at different heights align with each other azimuthally. The thermometry is in a similar layout on the aspect ratio one tank ($\Gamma = 1.0$). b) Photo of the convection cell of the RoMag device. c) Schematics of the convection cell of the RoMag device. For further device details, see \citet{xu2022thermoelectric}.}
   \label{fig:Schematics}
\end{figure}

%
%
Our experiments are conducted using UCLA's RoMag device \cite{king2012heat,cheng2015laboratory,vogt2018jump,grannan2022experimental,xu2022thermoelectric}. Figure \ref{fig:Schematics}a) - c) show schematics of the diagnostics and the apparatus. The container consists of two copper end blocks and a stainless steel sidewall. Two sets of sidewalls, $\Gamma = 1.0$ and $2.0$, have been used in this study to investigate MC heat transfer from $10^5 \lesssim Ra \lesssim 10^8$. An external solenoid generates a steady vertical magnetic field, $0 < |\boldsymbol{B}_0| < 800$ Gauss, with a vertical component that varies within $\pm 0.5$\% over the field volume \citep{king2015magnetostrophic}. The tank is placed at the center of the solenoid's bore. A non-inductively wound electrical resistance pad heats the bottom of the lower copper end block at a constant rate, $0 < P \lesssim 2000 \mathrm{W}$, and a thermostated water-cooled heat exchanger maintains a constant temperature at the top of the upper copper end block. This setup can reach up to $Ra \approx 10^9$ and $Ch = 3\times 10^5$ in the $\Gamma = 1.0$ tank.  

Twelve thermistors in total are placed inside the top and bottom boundaries about $28.9\ \mathrm{mm}$ radially inwards from the sidewall, as shown in figure \ref{fig:Schematics}a). These end-block thermistors are used to measure the heat transfer efficiency of the system, characterised by the Nusselt number, 
\begin{equation}
    Nu = \frac{q H}{\lambda \Delta T},
\end{equation}
where $q = 4 P/(\pi D^2)$ is the heat flux, $P$ is the heating power, and $\lambda = 31.4 \ \mathrm{W/(m \cdot K)}$ is the thermal conductivity of gallium \cite{aurnou2018rotating}. The Nusselt number describes the total to conductive heat transfer ratio across the fluid layer, and $Nu=1$ corresponds to the conductive state. The vertical temperature difference across the fluid layer, $\Delta T$, is indirectly controlled by the constant basal heat flux. Six thermistors are attached to the sidewall midplane to detect wall modes and any thermal imprints of the bulk fluid structures at the sidewall. 

We have also conducted direct numerical simulations (DNS) using the finite volume code \textsc{goldfish} \goldfish \citep{horn2015toroidal,shishkina2015thermal, shishkina2016thermal, horn2017prograde, horn2018regimes, horn2019rotating, horn2021tornado}. The nondimensional equations governing quasi-static, Oberbeck-Boussinesq \citep{oberbeck1879warmeleitung,boussinesq1903theorie} magnetoconvection  are:
\begin{align}
\hspace*{-.45em}
\vect{\nabla} \cdot \widetilde{\vec{u}} &= 0, \label{eq:NS1}\\
D_{\widetilde{t}} \widetilde{\vec{u}} &= - \vect{\nabla} \widetilde{p}  + \sqrt{\frac{Pr}{Ra}}\, \vect{\nabla}^2 \widetilde{\vec{u}}  + \sqrt{\frac{Pr}{Ra \, Ek^2}}\, \widetilde{\vec{u}} \times  \vec{\hat{e}_z} + \sqrt{ \frac{Ch^2 Pr }{Ra}}\,  \widetilde{\vec{j}}\times \vec{\hat{e}_z} + \widetilde{T} \vec{\hat{e}_z}, \label{eq:NS2} \\
D_{\widetilde{t}} \widetilde{T} &=  \sqrt{\frac{1}{Ra Pr}}\, \vect{\nabla}^2 \widetilde{T},\label{eq:NS3}\\[.7em]
&\hspace*{-3.14em} \left. 
{\begin{array}{ll}
    \hspace*{.2em} 
    \vect{\nabla} \cdot \widetilde{\vec{j}} &= 0 \\[1em]
    \hspace*{1.7em} \widetilde{\vec{j}} &= - \vec{\nabla} \widetilde{\Phi} +  (\widetilde{\vec{u}} \times \vec{\hat{e}_z})
\end{array}
} \right \} \; \vec{\nabla}^2 \widetilde{\Phi} = \vec{\nabla} \cdot (\widetilde{\vec{u}} \times \vec{\hat{e}_z}), 
    \label{eq:NS4}
\end{align}
where $\widetilde{\vec{u}}$ denotes nondimensional velocity, $\widetilde{T}$ the temperature, $\widetilde{p}$ the pressure, $\widetilde{\vec{j}}$ the current density, and $\widetilde{\Phi}$ the electrostatic potential. The scales used for the nondimensionalisation are the free-fall speed $U_{f \! f} = \sqrt{\alpha g \Delta T H}$ \citep{aurnou2020connections}, the temperature difference between top and bottom $\Delta T$, the reference pressure $\rho_0 U_{f \! f}^2$, the reference current density $\sigma B_0 U_{f \! f}$ and the reference potential $B_0 H U_{f \! f}$.

Our non-linear DNS solves these equations in a cylindrical domain $(r, \phi, z)$ with  $\Gamma = 2.0$. The sidewall is assumed to be perfectly thermally insulating, $\partial_r T\mid_{\text{r=R}} = 0$, and the top and bottom plates are isothermal with $T_t = -0.5$ and $T_b = 0.5$, respectively.  All boundaries are assumed to be impermeable and no-slip, $\vect{u}\hspace*{-3pt}\mid_{\text{wall}} = 0$, and electrically insulating, $\vect{j}\hspace*{-3pt}\mid_{\text{wall}} = 0$, i.e. the current forms closed loops inside the domain. The DNS control parameters are set to $Pr = 0.027$,  $Ch = 4.0\times 10^4$, and $Ra = \{1.5\times 10^5,\ 2.0\times 10^5,\ 3.0 \times 10^5,\ 4.0 \times 10^5,\ 7.0\times 10^5,\ 1.0\times 10^6,\ 1.5\times 10^6,\ 4.0\times 10^6 \}$. The numerical mesh resolution is $N_r \times N_\phi \times N_z = 240 \times 256 \times 240$. This choice of mesh was verified by running simulations at twice the resolution for the highest $Ra$ for a shorter time, indicating the grid independence of the solution.

%% file: sections/4_results.tex
%

\subsection{Comparing onset predictions}
\begin{figure}[ht]
   \centering
    \includegraphics[width=\textwidth]{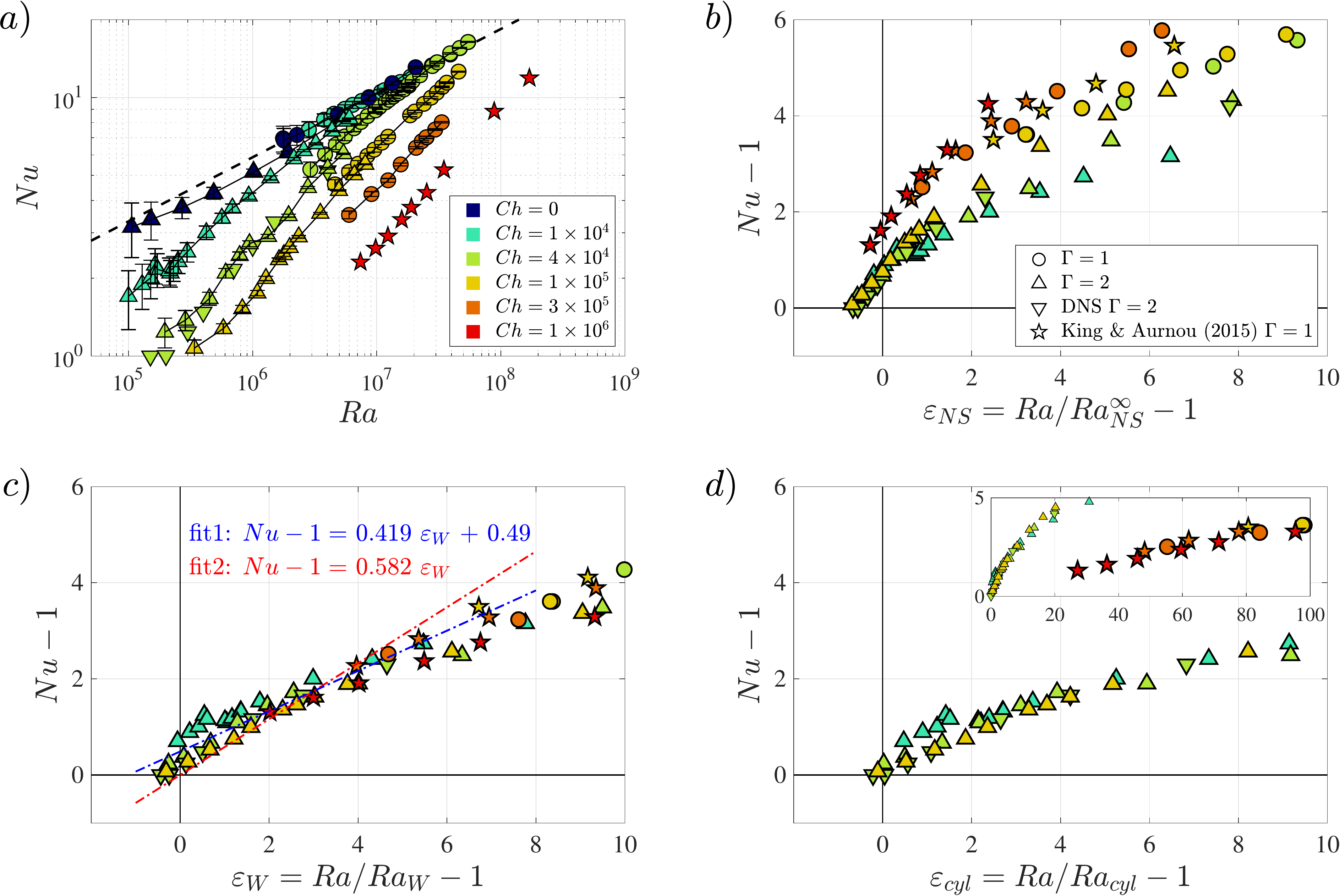}
 	\caption{a) $Nu - Ra$ survey with various $Ch$ from current study and one set of $Ch = 10^6$ data by \citet{king2015magnetostrophic}. Different colors correspond to different $Ch$ values. Different shapes correspond to different aspect ratios and sets of experiments, marked in the legend of panel b). Error bars are shown on the experimental data from the current study. The dashed line is the heat transfer scaling acquired from non-magnetic Rayleigh-B\'enard convection data in the $\Gamma = 1.0$ tank, marked by dark blue circles. See equation (\ref{eq:nurbc}a). Panel b)-d) show ratios of convective heat transfer to conduction ($Nu-1$) versus the reduced bifurcation parameter \citep{zhong1993rotating,horn2017prograde} using three different predicted critical Rayleigh numbers: b) infinite-plane stress-free critical $Ra$ defined in \citet{chandrasekhar1961hydrodynamic}; c) magnetowall mode critical $Ra$ by \citet{busse2008asymptotic}. d) \citet{houchens2002rayleigh}'s critical $Ra$ for magnetowall modes in two different aspect ratios, $\Gamma = 1$ and $2$. The linear fit in panel c) uses data close to onset at $\varepsilon_W \leq 5$. The $\Gamma = 1.0$ data in panel d) are shown in the smaller subplot, which only appears at $\varepsilon \gtrsim 25$.}
   \label{fig:RaRac2}
\end{figure}
The validity and accuracy of the onset predictions discussed in Section \ref{sec:2_theory} are tested here using both laboratory and DNS data at $10^5 \lesssim Ra \lesssim 10^8$, $0\lesssim Ch \lesssim 3\times 10^5$, and in $\Gamma = 1.0$ and $2.0$ cylindrical cells. 

Figure \ref{fig:RaRac2}a) shows measurements of heat transfer efficiency, $Nu$, as a function of the buoyancy forcing $Ra$. (All the detailed measurement data are provided in the tables in the Appendix). Vertical error bars based on heat loss and accuracy of the thermometry, are shown in the lab data from this study. We also include \citet{king2015magnetostrophic}'s $Ch=10^6$ data (red stars) made in the same $\Gamma = 1.0$ experimental setup used in this study. 
Figure \ref{fig:RaRac2}b) - d) show convective heat transfer data ($Nu-1$) as a function of supercriticality of the convection, as described by the reduced bifurcation parameter $\varepsilon = (Ra-Ra_{crit})/Ra_{crit}$, following the convention of \citep[][]{ecke1992hopf,zhong1993rotating,horn2017prograde}. Three different $Ra_{crit}$ are examined in panel b) to d): for convection in an infinite plane layer with two rigid boundaries, $Ra_{NS}^\infty$ (\ref{eq:ns}), wall-attached convection, $Ra_{W}^\infty$ (\ref{eq:Busse}), and convection in a cylinder with aspect ratio 1 and 2, $Ra_{cyl}$ (\ref{eq:cyl}). If the $Ra_{crit}$ prediction is accurate, the onset of convection occurs at $\varepsilon = 0$, and $Nu$ and follows an approximate linear scaling for sufficiently small $\varepsilon$. 

Figure \ref{fig:RaRac2}b) presents the convective heat transfer data, $Nu-1$, as a function of the reduced bifurcation parameter $\varepsilon_{NS} = (Ra-Ra_{NS}^\infty)/Ra_{NS}^\infty$ calcuated using eq.~\eqref{eq:ns}. The laboratory-numerical $Nu-1$ data exceeds 0 at $\varepsilon_{NS}<0$. This implies that the $Ra_{NS}^\infty$ predictions do not capture the onset of MC in our system. Moreover, the increased scatter and variation in $Nu$ for different $Ch$ as $\varepsilon_{NS}$ increases suggest a low correlation between the data and the expected linear $\varepsilon_{NS}$ scaling. As the infinite-plane $Ra_{crit}$ is associated with the bulk onset of convection, our heat transfer data implies that the MC flow does not initiate in the fluid bulk.

Figure \ref{fig:RaRac2}c) tests \citet{busse2008asymptotic}'s asymptotic onset predictions for magnetowall modes  as a function of $\varepsilon_W = (Ra-Ra_W)/Ra_W$ calculated using \eqref{eq:Busse}. The nonzero $Nu-1$ data start approximately at the origin of the graph, being only slightly below $\varepsilon_W = 0$ and show a good data collapse up to moderately high supercriticalities of  $\varepsilon \lesssim 10$. Thus, our data provide evidence that in our system the onset of convection occurs in the form of wall-attached modes. This is further quantified by two different linear least-square fits for the nonzero $Nu-1$ data for $\varepsilon_W \leq 5$. The first fit `fit1' (blue line) makes no assumptions on the onset and yields $Nu-1 = 0.419\ \varepsilon_W + 0.490$.  The second `fit2' (red line) is forced to pass through the origin, i.e. it assumes the onset prediction $\varepsilon_W = (Ra-Ra_W)/Ra_W$ is exact and yields $Nu-1 = 0.582\ \varepsilon_W$. Thus, in agreement with the theoretical predictions and as also found by \citet{zurner2020flow}, our MC onset data is consistent with the wall mode predictions. Furthermore, both linear fits hold well up to $\varepsilon_W \lesssim 10$, suggesting that the dynamics and the heat transfer in our system are largely controlled by linear magnetowall modes within this supercriticality range \citep[cf.][]{ecke1992hopf,zhong1993rotating,horn2017prograde,lu2021heat}. 

Figure \ref{fig:RaRac2}d) tests the hybrid theoretical-numerical predictions of \citet{houchens2002rayleigh} for magnetowall modes in cylindrical geometries by plotting $Nu - 1$ versus $\varepsilon_{cyl} = (Ra-Ra_{cyl})/Ra_{cyl}$ calculated using \eqref{eq:cyl}. The underlying assumptions for these predictions best match the experimental and numerical setup. Therefore, they should best capture the measured onset of convection. The $\Gamma = 2.0$ data (green and yellow hues) show an excellent agreement with theory. 
The $\Gamma = 1.0$ data (inset, red and orange hues) does not have a low enough supercriticality ($\varepsilon_{cyl} > 28$) to reliably test the exact $Ra_{cyl}$ value. Thus, we are currently unable to disambiguate which magnetowall mode onset predictions are more accurate, even though \citet{houchens2002rayleigh}'s $\Gamma = 1.0$ predictions differ from \citet{busse2008asymptotic}'s by nearly an order of magnitude.
\begin{figure}[b!]
    \centering
    \includegraphics[width=\textwidth]{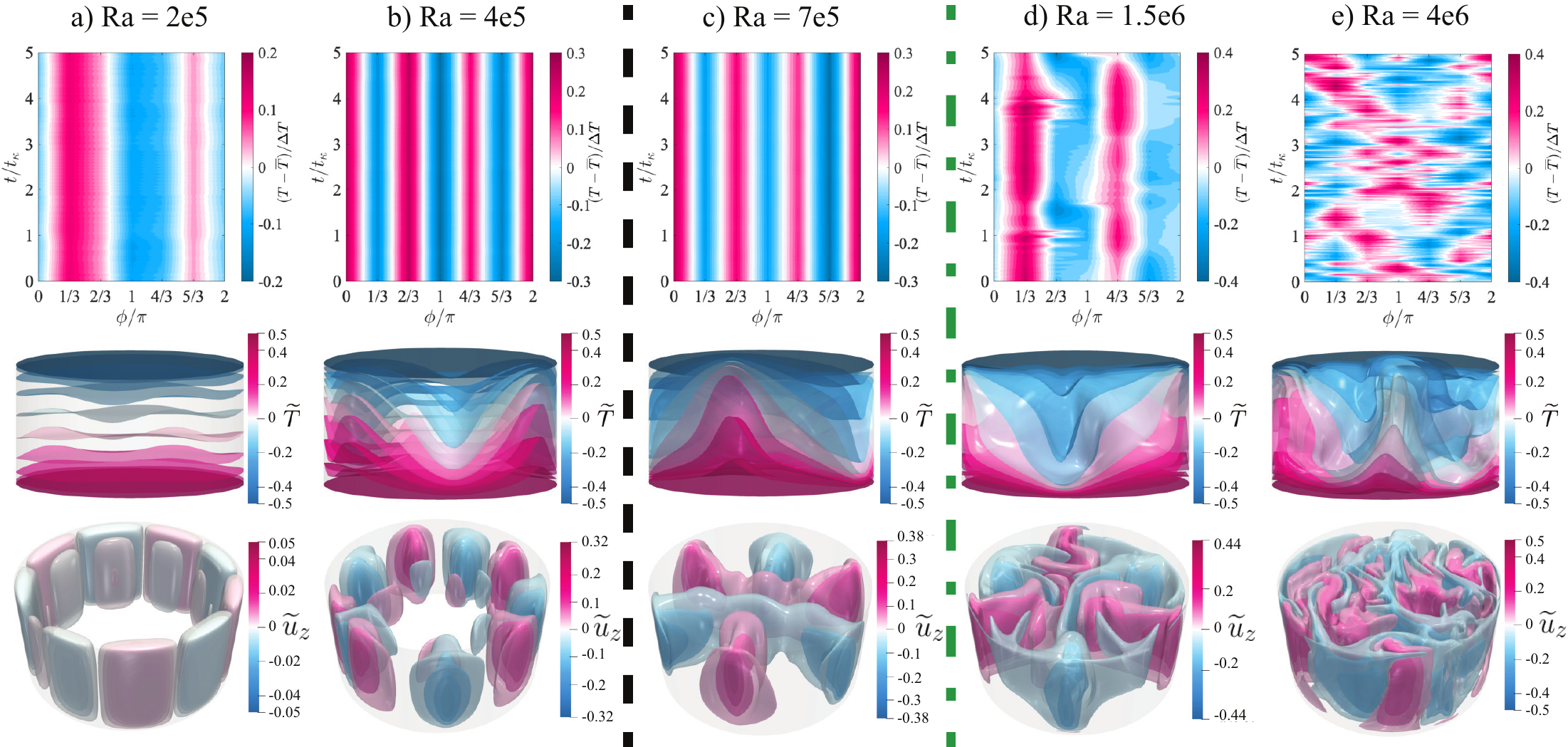}
    \caption{Temperature and velocity fields from $\Gamma = 2.0$ laboratory experiments at $Ch = 4.0\times 10^4$. The vertical columns show cases at a) $Ra = 2.0\times 10^5$, b) $Ra = 4.0\times 10^5$, c) $Ra = 7.0\times 10^5$, d) $Ra = 1.5\times 10^6$, and e) $Ra = 4.0\times 10^6$, respectively ($Ra$ is only approximate for the laboratory cases; their exact values are given in tables \ref{table1} and \ref{table1b}). The first row shows the azimuthal-temporal temperature contours at the midplane interpolated by lab data over $5$ thermal diffusion times, $\tau_\kappa = H^2/\kappa$. The color represents the dimensionless temperature, $(T-\overline T)/\Delta T$, where $\overline T$ is the mean temperature obtained by averaging the top and bottom temperatures. The second row consists of snapshots of the normalised DNS temperature field $\widetilde {T}$ and the third row presents snapshots of normalised DNS vertical velocity fields $\widetilde{u}_z$ at the same moment in time as the temperature field. The vertical black dashed line between b) and c) separates between cases below bulk onset (based on $Ra_{NS}^{\infty}$) to the left and above bulk onset to the right. The vertical green dash-dotted line between c) and d) indicates the transition from an azimuthal mode number of $m=3$ to $m\leq 2$ seen in the laboratory cases.}
   \label{fig:multipanel}
\end{figure}

\subsection{Transition to multimodality}
\begin{figure}[b!]
   \centering
    \includegraphics[width=\textwidth]{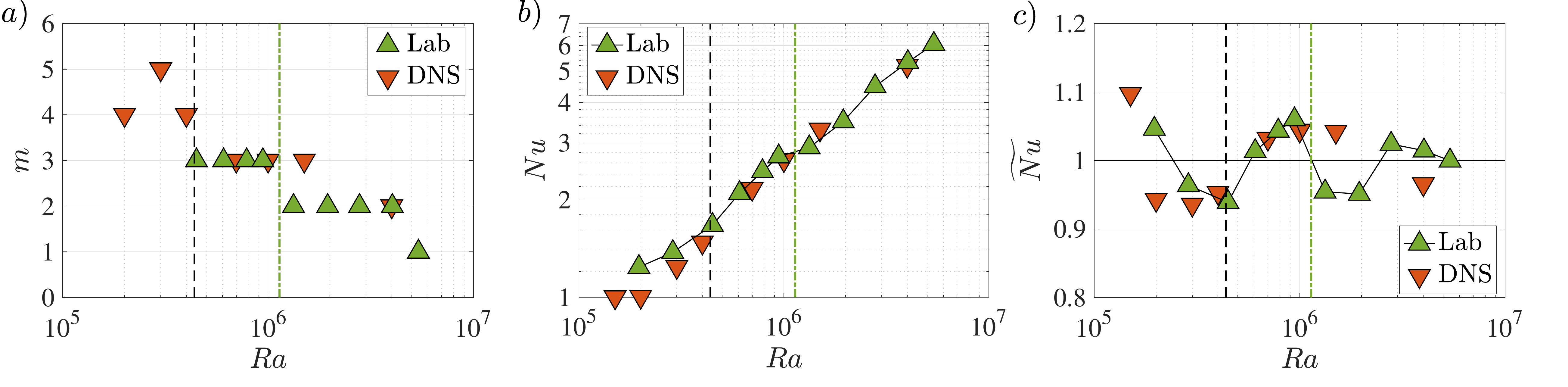}
    \caption{a) Azimuthal mode number $m$ as a function of $Ra$ at $Ch = 4\times 10^4$, $\Gamma = 2.0$ case for both lab (green triangles) and numerical data (red triangles). b) Nusselt number $Nu$ as a function of $Ra$ for both laboratory experimental and numerical data. c) $Nu$-$Ra$ Data plotted on detrended curves. Here, $\widetilde{Nu}$ is $Nu$ normalised by the linear fits (\ref{eq:detrendlab}, \ref{eq:detrenddns}) of each respective data set in panel b). In parallel to figure \ref{fig:multipanel}, the black dashed line between b) and c) indicates the predicted $Ra_{NS}^{\infty}$, whereas the green dash-dotted line between c) and d) marks the average $Ra$ between two adjacent laboratory data with $m=3$ and $m=2$. The kinks in $Nu$-$Ra$ data in b) are shown as the fluctuations around $\widetilde{Nu} = 1$ in the $\widetilde{Nu}-Ra$ trend, likely associated with mode switching.}
   \label{fig:modenumber}
\end{figure}
We analysed temperature and velocity field data to elucidate both the wall modes and the transition to multi-modal flow at higher supercriticality. Figure \ref{fig:multipanel} shows temperature and velocity fields of five laboratory numerical cases with $Ch \simeq 4\times 10^4$ in the $\Gamma = 2.0$ tank. Each column represents the laboratory case (top row) and its corresponding numerical case (middle and bottom rows) at a similar $Ra$. The detailed parameters are given in table \ref{table1} and \ref{table1b} in the appendix. The top rows show nondimensional temperature $(T-\overline T)/\Delta T$ Hovm\"oller diagrams, a time evolution of the sidewall midplane temperature field. The temperature fields $T$ are interpolated by the mid-plane thermistor data taken $60^\circ$ apart in azimuth. The mean temperature $\overline T$ is measured by averaging the top and bottom boundaries' temperature; the vertical temperature difference across the fluid layer is denoted as $\Delta T$. The second and third rows of figure \ref{fig:multipanel} show snapshots of numerical 3D isosurfaces of the dimensionless temperature fields $\widetilde T = (T-\left<T\right>)/\Delta T$ and corresponding vertical velocity fields $\widetilde{\boldsymbol{u}}_z$ at the same instant in the time. The mean temperature $\left<T\right>$ is calculated by averaging the temperature fields over the entire domain. 

The velocity and temperature fields in figure \ref{fig:multipanel} all show magnetowall modes, manifesting as azimuthally alternating upwelling warm and downwelling cold patches located close to the sidewall. The snapshots of the DNS velocity fields in figures \ref{fig:multipanel}a) and b) further reveal that the magnetowall modes have a two-layer, `nose-like' flow pattern attached to the sidewall with alternating $\pm \widetilde{\boldsymbol{u}}_z$. \citet{liu2018wall} observed similar structures in their simulations in a rectangular box at $Ra = 10^7,\ Ch = 4\times 10^6$. They found that these noses scale approximately with a Shercliff boundary layer thickness $\delta_{Sh} \propto Ch^{-1/4}$ \citep{liu2018wall}. 

Figure \ref{fig:multipanel}a) to b) show that these noses also grow gradually towards the interior as the supercriticality increases, while the interior remains otherwise quiescent. This ``Pinocchio effect"  persist until $Ra \gtrsim Ra_{NS}^\infty$, when the bulk fluid starts convecting from the top and bottom boundaries and then interacts with the inward-extended wall modes. 

Figure \ref{fig:multipanel}c) shows this extending nose behavior for $Ra = 7\times 10^5$, which is just above the bulk onset $Ra > Ra_{NS}^\infty$. The DNS velocity field visualises how two noses with positive/negative $u_z$ (pink/blue) connect across the entire diameter of the tank via the convecting upwelling/downwelling fluid in the interior. The laboratory and numerical temperature field on the sidewall agree perfectly and show that close to the sidewall the wall modes are virtually unaffected by this interior dynamics.
In total there are six alternating cold and hot patches along the sidewall azimuth, i.e., the azimuthal mode number is $m=3$. The magnetowall mode number $m$ is defined as the number of repeating azimuthal structures along the lateral surface.

Figure \ref{fig:multipanel}d) and e) show that these nonlinear interactions become more complicated and chaotic as $Ra$ increases further. The nose-like structures interact and impinge on each other. The nonlinear behaviour also affects the flow close to the sidewall as visible in the temperature Hovm\"oller diagram from sidewall thermometry in d) for $Ra = 1.5\times 10^6$ and even more so in e) for $Ra = 4.0\times 10^6$.

For $Ra = 4.0\times 10^6$ (Figure \ref{fig:multipanel}e), the experimental temperature Hovm\"oller diagram shows that the magnetowall modes are transient between $m=1$ and $m=2$ in a chaotic sequence. The velocity field of the DNS further demonstrates that the bulk flow dominates the dynamics, and, hence the flow for $Ra = 4.0\times 10^6$ significantly differs from the ones at lower $Ra$ shown in figure \ref{fig:multipanel}a) to d). 

There is, however, a small discrepancy between the number of azimuthal structures between the lab and DNS data, being $m = 2$ and $m = 3$, respectively for $Ra = 1.5\times 10^6$ (Figure \ref{fig:multipanel}d). This may be because $m$ is sensitive to small changes in $Ra$ and $\Gamma$, and there are slight differences in parameters between the lab and the DNS, or because of the sidewall boundary conditions which are not perfectly adiabiatic in the experiment. It is also possible that the DNS snapshots do not capture fully equilibrated flow patterns whilst the lab experiments revealed more averaged dynamics of MC, as, unlike the DNS, they can be run for many thermal diffusion times.

Figure \ref{fig:modenumber}a) shows how the time-averaged azimuthal mode numbers $m$ observed in the laboratory experiment and the DNS velocity fields depend on the Rayleigh number $Ra$. The $m$ values generally decrease with increasing $Ra$, which qualitatively agrees with previous studies \citep{liu2018wall,akhmedagaev2020turbulent,zurner2020flow}. For $Ra < 10^4$, we only present DNS data in this plot and no lab data due to a combination of both precision and spatial aliasing issues of the sidewall thermistor array. The temperature variation between each wall mode structure near the midplane is $\lesssim 0.2\ \mathrm{K}$, which is too small to be resolved by our thermometers. Additionally, with only six thermistors evenly spaced at the azimuth, we can only resolve up to $m = 3$ according to the Nyquist-Shannon sampling theorem. Thus, even though the first-row temperature contour in figure \ref{fig:modenumber}b) shows an $m = 3$ structure, it was omitted in figure \ref{fig:modenumber}a) and only the $m = 4$ from the velocity field from the DNS is shown.

The changes in mode number also affect the global heat transport. Figure \ref{fig:modenumber}b) shows that $Nu$ increases monotonically with $Ra$, but not at a constant rate. Instead, kinks exist in the $Nu$-$Ra$ trends in both lab experiments and DNS for $Ch = 4\times 10^4$ (and $Ch = 10^5$, see figure \ref{fig:RaRac2}a), a phenomenon which has not been reported in previous MC experiments \citep{cioni2000effect, zurner2020flow}. To further investigate this behaviour, we normalised $Nu$ by power laws obtained by separate fits to the $Ch = 4 \times 10^4$ laboratory and DNS $Nu$-$Ra$ data sets.  For the laboratory data, the best fit is
    \begin{equation}
       \widetilde{Nu} = Nu/(0.0029Ra^{0.493}),
     \label{eq:detrendlab}
    \end{equation}
    whereas for the DNS, it is found that
    \begin{equation}
        \widetilde{Nu} = Nu/(0.0014Ra^{0.541}).
        \label{eq:detrenddns}
    \end{equation}

Figure \ref{fig:modenumber}c) shows the normalised $\widetilde{Nu}$. The non-monotonicity of the trend manifests as fluctuations around $\widetilde{Nu} \approx 1$ with an amplitude of approximately $0.05$. The increase after the first local minimum in the experimental $\widetilde{Nu}$ data curve coincides with the bulk onset, $Ra = Ra_{NS}^{\infty}$, and is marked by the vertical black dashed line. This suggests that bulk convection enhances heat transfer efficiency. The decrease after the first local maximum in the experimental $\widetilde{Nu}$ data curve coincides with the change of mode numbers from $m=3$ to $m=2$ observed in the laboratory cases (cf. figure \ref{fig:multipanel}) and is marked by the green dash-dotted line. The transition to a smaller mode number appears to suppress the heat transfer efficiency temporarily. A similar behaviour was observed in \citet{horanyi1999turbulent}'s liquid metal Rayleigh-B\'enard convection experiments. The second enhancement in $\widetilde{Nu}$ after the second local minimum happens when the highly-nonlinear flow structures in the bulk fluid start to dominate the convective dynamics. This corresponds to flow behaviors somewhere between $Ra = 1.5 \times 10^6$ (figure \ref{fig:multipanel}d) and $Ra = 4 \times 10^6$ (figure \ref{fig:multipanel}e). The DNS data in figure \ref{fig:modenumber}c) match the first enhancement near $Ra = Ra_{NS}^{\infty}$. Because no mode switch  from $m = 3$ to $m = 2$ was found in the DNS, no kink shows up in the  $Nu$-$Ra$ trend in the DNS data at this point. 

%% file: sections/5_discussion.tex
%
%
\subsection{Wall modes stability and the cellular flow regime}

In our laboratory-numerical experiments, the magnetowall modes are stationary and do not drift over dynamically long time scales ($\gg 5 \tau_\kappa$), in contrast to the drifting wall modes in rotating convection systems \citep[e.g.,][]{ecke1992hopf}. This is because the quasi-static Lorentz force ($f_L \propto B_0^2$) does not break the system's azimuthal symmetry, unlike the Coriolis force \citep{ecke1992hopf}). The stationarity of magnetowall modes has been confirmed in both numerical simulations \citep{liu2018wall} and laboratory experiments \citep{zurner2020flow}. Furthermore, \citet{liu2018wall} showed that the magnetowall modes can inject jets into the bulk. This phenomenon was also found in the numerical MC simulations of \citet{akhmedagaev2020turbulent}, where strong, axially invariant wall mode injections were accompanied by a net azimuthal drift of the flow field with random orientations. We believe that the collisional interaction of the jets in a small aspect ratio cylinder ($\Gamma = 1.0$), rather than any innate azimuthal motion of magnetowall modes, is responsible for the drifting motions observed by \citet{akhmedagaev2020turbulent}. 

The fully three-dimensional flow fields from our DNS facilitated the investigation of the bulk flow patterns in this study. Thus, we are also able to compare multiple cases at similar parameters with \citet{zurner2020flow} who inferred the interior structure solely from linewise Ultrasonic Doppler Velocimetry (UDV) and pointwise temperature measurements along the sidewalls and within the top and bottom plate. Our identified flow structures and corresponding flow changes match well with their observations. Specifically, what they denoted as the 'cellular regime' corresponds to our case with extended wall mode noses with interior bulk modes. Our $Ra = 7 \times 10^5$ and $Ra = 1.5 \times 10^6$ cases (figure \ref{fig:multipanel}c, d) resemble the inferred '3-cell' and '4-cell' patterns of figure 3 in \citet{zurner2020flow}. Our $Ra = 1.5\times 10^6,\ m=2$ thermal data in figure \ref{fig:multipanel}d) also agrees with their '2 cell' pattern on the sidewall. Moreover, the transition range from the 'cellular regime' to the non-rotating LSC regime in their experiment occurred approximately at $Ra \gtrsim 4\times 10^6$ for $Ch = 4 \approx 10^4$, which is consistent with our observation of a more chaotic interior and unsteady and irregular wall mode behaviour, as shown in figure \ref{fig:multipanel}e).

\subsection{The $Nu$ vs. $Ra$ MC party}
\begin{figure}[t!]
   \centering
    \makebox[\textwidth][c] {\includegraphics[width=\columnwidth]{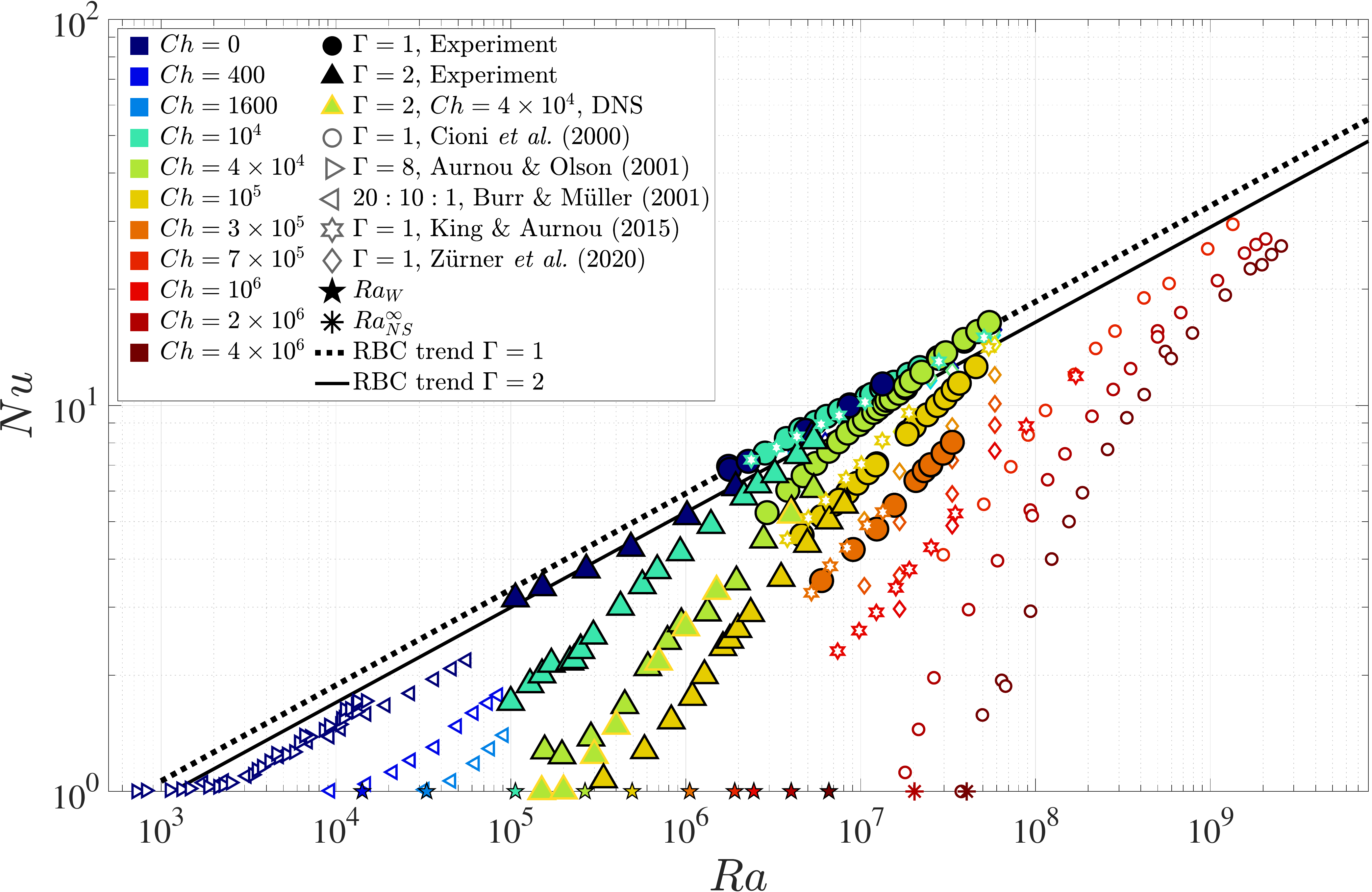}} 
     	  \caption{Collection of $Nu$-$Ra$ data from this study and previous MC laboratory experiments in liquid metal \cite{chandrasekhar1961hydrodynamic,cioni2000effect,aurnou2001experiments,burr2001rayleigh,king2015magnetostrophic,zurner2020flow}. Color represents $Ch$. The filled symbols mark the data in this study. The five-pointed stars at $Nu = 1$ mark the $Ra_W$ (\ref{eq:Busse}) for all different $Ch>0$. The two asterisk symbols from left to right mark $Ra^{\infty}_{NS}$ for $Ch=2\times 10^6$ and $4\times 10^6$, respectively, and corresponding to the \citet{cioni2000effect}'s two $Ch$ data set. The non-filled color symbols are selected heat transfer data from prior liquid metal laboratory experiments. All data displayed here are included in appendix tables \ref{table1} - \ref{table7}.}
   \label{fig:NuRaParty}
\end{figure}

Figure \ref{fig:NuRaParty} presents a broad compilation of laboratory MC heat transfer measurements in different aspect ratios and geometries, all in the presence of an external vertical magnetic field. 
\citet{cioni2000effect} (open circles) studied liquid mercury in a $\Gamma = 1.0$ cylindrical cell up to $Ch \approx 4\times 10^6$ and $Ra \approx 3\times 10^9$. \citet{aurnou2001experiments} (open triangles pointing right) carried out near onset liquid gallium experiments in a $\Gamma = 8$ rectangular cell. \citet{burr2001rayleigh} (open triangles pointing left) investigated sodium-potassium alloy in a $20:10:1$ rectangular cell. \citet{king2015magnetostrophic} studied liquid gallium MC in a $\Gamma = 1.0$ cylinder on the same device (RoMag) as this study. \citet{zurner2020flow} studied both heat and momentum transfer behaviors of liquid GaInSn in a $\Gamma = 1$ cylinder. The results from our current $\Gamma = \{1.0,\ 2.0\}$ cylindrical liquid gallium experiments and simulations are demarcated by the filled symbols. 

In addition, we have included the Nusselt number data for $Ch=0$, i.e. pure Rayleigh-B\'enard convection (RBC). Best fits to the RBC cases yield
\begin{subequations}
\begin{eqnarray}
    Nu_{0} &\approx (0.191 \pm 0.088) Ra^{0.248 \pm 0.025} \quad \text{for}\ \Gamma = 1.0,  \label{eq:nurbca} \\
    Nu_{0} &\approx (0.176 \pm 0.081) Ra^{0.246 \pm 0.028} \quad \text{for}\ \Gamma = 2.0,  \label{eq:nurbcb}
\end{eqnarray}
    \label{eq:nurbc} 
\end{subequations}
\hspace{-3pt}which are in good agreement with previous studies on the same device (RoMag) \citep{king2013flow,king2015magnetostrophic,vogt2018jump,aurnou2018rotating}. The differences between these two scaling laws lie within their error bars but may be due to the different tank aspect ratios \citep{king2013turbulent,vogt2018jump,aurnou2018rotating}.  

As discussed in section \ref{sec:4_results}, our data show that the onset of MC in a cylinder occurs via wall modes. The five-point stars at $Nu = 1$ mark the magnetowall mode onset predictions $Ra_W$ \eqref{eq:Busse} for the different $Ch$. Our near-onset data at $Ch = 10^5$ (yellow triangles) is in good agreement with the onset prediction by \citet{busse2008asymptotic} (yellow star). However, \citet{cioni2000effect}'s heat transfer data at $Ch = 2\times 10^6$ and $Ch = 4\times 10^6$ have $Ra_{crit} \approx 3 Ra_W$. The lowest $Nu$ data from \citet{cioni2000effect}'s $Ch = 2\times 10^6$ and $Ch = 4\times 10^6$ align well with the bulk onset prediction, $Ra^{\infty}_{NS}$. \citet{zurner2020flow} have analysed the $Nu$-$Ra$ trends and also found a large deviation between the experimental results of \citet{cioni2000effect} and \citet{king2015magnetostrophic}. This discrepancy is likely due to \citet{cioni2000effect}'s thermometry setup, which used a single thermistor at the center of each top and bottom boundary to measure $\Delta T$. This setup was not designed to characterise wall modes and could only detect the convective heat transfer occurring near the center of the tank.  Thus, top and bottom end wall temperature measurements nearer to the sidewall are required in order to detect the onset of wall modes and to measure their contributions to  the total heat transfer \citep[cf.][]{akhmedagaev2020turbulent, zurner2020flow, grannan2022experimental}.

\subsection{Comparison between magnetoconvection and rotating convection in liquid metal}
The goal of this work is to provide a better understanding of the pathway from convective onset to multimodal turbulence in liquid metal magnetoconvection.  Thus far, we have compared our laboratory-numerical data with the results of other MC studies. Here we expand on this by comparing our MC data against rotating convection data. Although the Lorentz and Coriolis forces both act to constrain the convection in these systems \citep{julien2007reduced}, their data are rarely closely compared since the vast majority of rotating convection (RC) studies are carried out in moderate to high Prandtl fluids (non-metals), whereas MC studies are nearly always made using low $Pr$ liquid metals \citep[cf.][]{Aujogue18}.

\begin{figure}[b!]
   \centering
    \includegraphics[width=\textwidth]{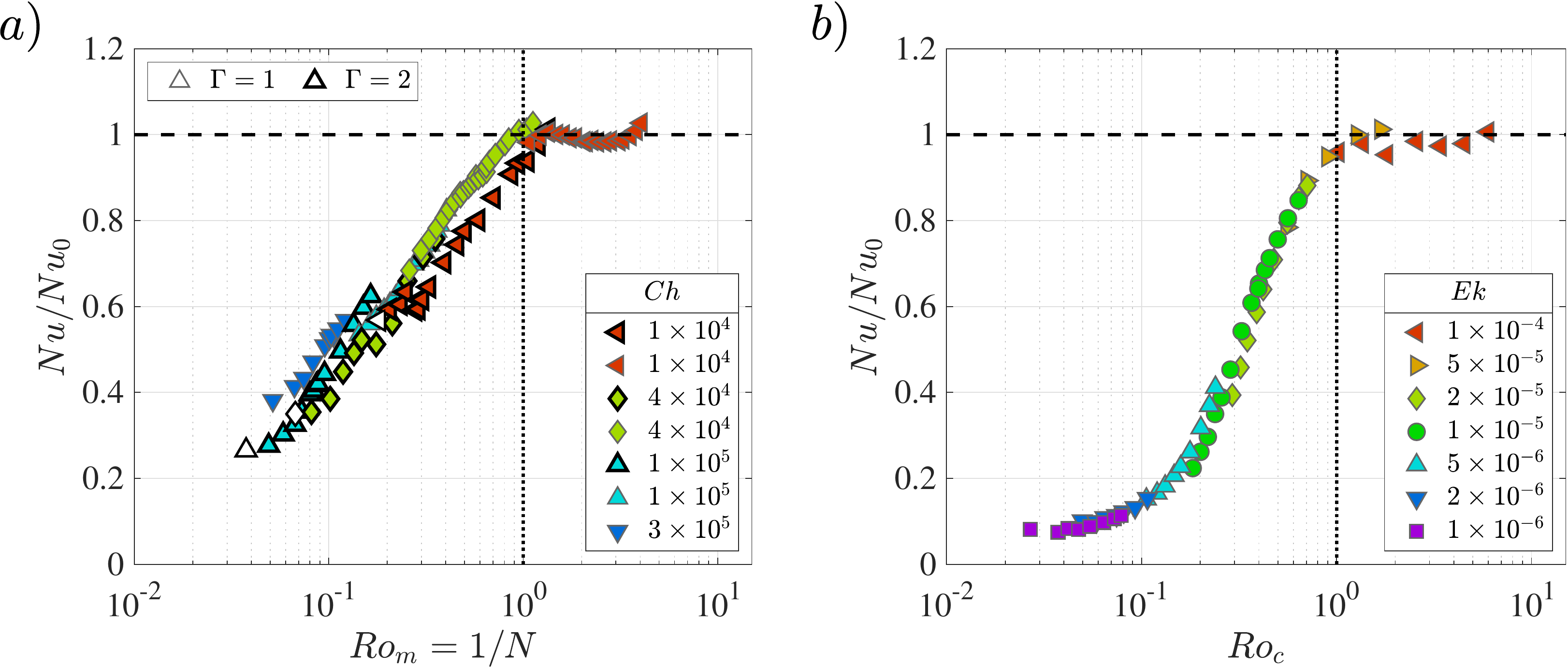}
 	  \caption{
 	  a) Normalised Nusselt number ($Nu/Nu_{0}$) versus magnetic Rossby number ($Ro_m$) for MC lab data in $\Gamma = 1.0$ and $2.0$ tanks. Here, $Nu_{0}$ are the best-fit power laws in (\ref{eq:nurbc}) and $Ro_m$ is the inverse of interaction parameter $N$, as defined in (\ref{eq:Rom}). Symbols in thick black outlines represent $\Gamma = 2.0$ data, and those in thin grey outlines are $\Gamma=1.0$ data. The color of the symbols indicates $\mathrm{log_{10}}(Ch)$. The white symbols are subcritical cases according to wall mode onset ($Ra<Ra_{W}$). The symbol shapes do not contain information but help differentiate different $Ch$. b) Rotating convective heat transfer data adapted from the $\Gamma = 1.0$ liquid gallium experiments of \citet{king2013turbulent}. The color indicates $\mathrm{log_{10}}(Ek^{-1})$. The vertical axis shows a reduced Nusselt number $Nu/Nu_{0}= Nu/(0.185 Ra^{0.25})$, following \citep{king2013turbulent}. The horizontal axis is convective Rossby number $Ro_c$, as defined in (\ref{eq:Rom}).} 
   \label{fig:MCvsRC}
\end{figure}

Figure \ref{fig:MCvsRC} shows a side-by-side comparison of the convective heat transfer efficiency $Nu/Nu_0$ as a function of the normalised buoyancy forcing in a) our present liquid gallium $\Gamma = 1.0$ and $2.0$ MC experiments and b) the liquid gallium $\Gamma = 1.0$ rotating convection data from King \& Aurnou (2013) \cite{king2013turbulent}. The thicker outlined symbols in panel a) demarcate the $\Gamma = 2.0$ MC cases. The liquid gallium convection data in Figure \ref{fig:MCvsRC} was all obtained using the same experimental apparatus and setup. The fill color in a) denotes $\log_{10}(Ch)$, whereas it denotes $\log_{10}(Ek^{-1})$ in panel b). The Ekman number, $Ek = \nu / 2 \Omega H^2$, is the ratio of viscous and Coriolis forces in rotating systems and $\Omega$ is the RC system's angular rotation rate.  

The best co-collapse of the $Nu/Nu_0$ data sets was found when the buoyancy force was normalised by the appropriate constraining force, that being Lorentz in MC and Coriolis forces in RC.  In MC, this non-dimensional ratio is called the magnetic Rossby number, $Ro_m$, which formally describes the ratio of convective inertia and the Lorentz force: 
\begin{equation}
    Ro_m= \frac{\text{Inertia}}{\text{Lorentz}} = Re_{f\!f}Ch^{-1}= \sqrt{\frac{RaCh^{-2}}{Pr}},
    \label{eq:Rom}
\end{equation}
where $Re_{f\!f} = U_{f\!f}H/\nu$. In the MHD literature, the reciprocal of this ratio, which is called the interaction parameter $N = Ro_m^{-1}$ is often employed \citep[e.g.,][]{xu2022thermoelectric}.  In RC, this non-dimensional ratio is called the convective Rossby number, $Ro_c$, 
\begin{equation}
    Ro_c= \frac{\text{Inertia}}{\text{Coriolis}} = Re_{f\!f}Ek = \sqrt{\frac{RaEk^2}{Pr}},
    \label{eq:Roc}
\end{equation}
which shows up as a collapse parameter in a broad array of rotating convection problems \citep[e.g.][]{gastine2014zonal,aurnou2020connections, Landin23}. Lorentz forces dominate in MC when $Ro_m = 1/N$ is small; Coriolis forces dominate in RC when $Ro_c$ is small. When these Rossby numbers exceed unity, buoyancy-driven inertial forces should be dominant and the convection is expected to be effectively unconstrained on all available length scales in the system. 

Comparing Figures \ref{fig:MCvsRC}a) and \ref{fig:MCvsRC}b), it is clear that the liquid metal MC and RC data have similar gross morphologies.  The $Nu/Nu_0$ is near unity and effectively flat for both $Ro_m \gtrsim 1$ and $Ro_c \gtrsim 1$. Thus, when the constraining Lorentz or Coriolis forces become subdominant to inertia in either system, the heat transfer is similar to that found in unconstrained RBC experiments. 

The basic structures of MC and RC data are also similar at $Ro_m \lesssim 1$ and $Ro_c \lesssim 1$: the normalised heat transfer trends relatively sharply downwards with decreasing Rossby number.  However, the detailed structures of the low $Ro_m$ and low $Ro_c$ data differ substantively. The data fall off more steeply with Rossby in the rotating case, then it greatly flattens out in the lowest $Nu/Nu_0$ RC cases. The differences in slope may be due to the difference in Ekman pumping (EP) effects in both systems \citep{julien2016nonlinear}, although heat transfer enhancement by EP is typically weak in metals since it is hard to modify the thermal boundary layers in low $Pr$ flows. 

Alternatively, these differences may be caused by the differences in critical $Ra$ values and their scalings. For instance, in the parameter ranges explored in Figure \ref{fig:MCvsRC}, oscillatory bulk convection first onsets in RC \citep{aurnou2018rotating}, whereas it is the wall modes that develop first in MC.  Further, the bulk magnetoconvective onset scales asymptotically as $Ra_{crit} \sim Ch^1$ whereas bulk oscillatory convective onset asymptotically scales as $Ra_{crit} \sim Ek^{-4/3}$.  This $1/3$ difference in the scaling exponents may imply that the available range of $Nu/Nu_0$ will be larger in the RC cases. Further, the flat tail in the lowest $Nu/Nu_0$ RC data is likely due to the low convective heat transfer efficiency of oscillatory rotating convection.

Thus, the gross structures of the two data compilations are similar in Figure \ref{fig:MCvsRC}. We hypothesise that their differences in our current data are likely due to the various modal onset phenomena, as are clearly present in Figure \ref{fig:modenumber}, that alter the low Rossby branches of each figure panel. However, it may be that differences in MC and RC supercritical dynamics better explain these data \citep[cf.][]{yan2019heat, Oliver23}. Regardless of the root cause, these low Rossby differences have thus far thwarted our attempts to create a unified plot in which all the $Nu/Nu_0$ data are simultaneously collapsed \citep[cf.][]{Chong17}. 

\subsection{Summary}

We have conducted a suite of laboratory thermal measurements of liquid gallium magnetoconvection in cylindrical containers of aspect ratios $\Gamma = 1.0$ and $2.0$. Our data allow us to characterise liquid metal MC from wall mode onset to multimodality. We performed a fixed $Ch = 4\times 10^4$ survey of direct numerical simulation for the same system in a $\Gamma = 2.0$ cylindrical geometry. Both laboratory and numerical methods obtained similar heat transfer behaviors, with possible subtle differences in flow morphology. Together with previous studies, our liquid metal heat transfer data comprise a convective heat transfer survey over six orders of magnitude in both $Ra$ and in $Ch$ (Figure \ref{fig:NuRaParty}).

\citet{busse2008asymptotic}'s asymptotic solutions for magnetowall modes best collapse all our MC heat transfer data, whereas the hybrid theoretical-numerical solutions by \citet{houchens2002rayleigh} captures the exact onset for $\Gamma = 1.0$, but the onset for $\Gamma = 2.0$ remains unverified. Better theoretical onset predictions are needed for liquid metal MC in a cylindrical cell as a function of $\Gamma$. This differs from liquid metal rotating convection where accurate theoretical predictions currently exist for low-$Pr$ fluids in cylindrical geometries \citep{zhang2009onset, zhang_liao_2017}. 

The MC flow morphology was characterised experimentally using a sidewall thermistor array as well as the DNS temperature and velocity fields. The onset of convection was verified to occur in the form of stationary (non-drifting) magnetowall modes. These magnetowall modes develop nose-like protuberances that extend into the fluid bulk with increasing supercriticality. At Rayleigh numbers beyond the critical value for steady bulk convection, the noses interact with the interior bulk modes, likely resulting in the apparent cell-like flow patterns observed by \citet{zurner2020flow}. Our data show that MC convective heat transport is sensitive to the flow morphology, with the Nusselt number $Nu$-$Ra$ data containing distinct kinks at these points where the dominant convection mode appears to change. 

Lastly, liquid metal heat transfer trends in magnetoconvection were compared with rotating convection. The gross behavior of the heat transfer is controlled by the magnetic and convective Rossby numbers, $Ro_m$ and $Ro_c$, in the respective systems, with the normalised heat transport $Nu/Nu_0$ approaching the RBC scaling as $Ro_m$ and $Ro_c$ approach unity from below. The detailed trends at Rossby values less than unity show clear differences between MC and RC.  We have not yet deduced a scheme by which it is possible to collapse all the liquid metal MC and RC data in a unified way.  

%% file: sections/appendix.tex
%
%

\begin{center}
\begin{table}[ht]
\begin{ruledtabular}
\begin{tabular}{ ccc |ccc |ccc }
$ \bm{Ch}$  & $\bm{Ra}$ & $\bm{Nu}$  & $ \bm{Ch}$  & $\bm{Ra}$ & $\bm{Nu}$ & $ \bm{Ch}$  & $\bm{Ra}$ & $\bm{Nu}$ \\
\hline\\[-2.5ex]
0        & 1.93E+06 & 6.13 & 1.00E+04 & 5.64E+05 & 3.41 & 3.98E+04 & 1.93E+06 & 3.52 \\
0        & 1.02E+06 & 5.16 & 1.01E+04 & 6.87E+05 & 3.74 & 4.00E+04 & 2.79E+06 & 4.48 \\
0        & 4.86E+05 & 4.27 & 1.00E+04 & 9.30E+05 & 4.16 & 4.02E+04 & 4.03E+06 & 5.33 \\
0        & 2.70E+05 & 3.75 & 1.02E+04 & 1.39E+06 & 4.89 & 4.15E+04 & 5.40E+06 & 6.06 \\
0        & 1.52E+05 & 3.37 & 1.03E+04 & 2.15E+06 & 5.79 & 9.40E+04 & 1.65E+05 & 1.20 \\
0        & 1.06E+05 & 3.16 & 1.04E+04 & 2.58E+06 & 6.23 & 9.40E+04 & 2.10E+05 & 1.15 \\
0        & 3.81E+04 & 3.24 & 1.04E+04 & 3.23E+06 & 6.62 & 9.42E+04 & 3.39E+05 & 1.07 \\
0        & 5.69E+04 & 3.99 & 1.06E+04 & 4.37E+06 & 7.42 & 9.44E+04 & 5.81E+05 & 1.28 \\
0        & 5.26E+06 & 8.31 & 1.05E+04 & 5.40E+06 & 8.11 & 9.49E+04 & 8.26E+05 & 1.52 \\
1.06E+04 & 9.98E+04 & 1.70 & 1.09E+04 & 6.01E+06 & 8.28 & 9.47E+04 & 1.09E+06 & 1.75 \\
1.06E+04 & 1.29E+05 & 1.89 & 3.87E+04 & 9.24E+04 & 1.47 & 9.48E+04 & 1.28E+06 & 2.00 \\
1.01E+04 & 1.51E+05 & 2.01 & 3.72E+04 & 1.56E+05 & 1.27 & 9.70E+04 & 1.63E+06 & 2.36 \\
1.00E+04 & 1.64E+05 & 2.25 & 3.88E+04 & 1.96E+05 & 1.24 & 9.73E+04 & 1.79E+06 & 2.47 \\
1.01E+04 & 1.70E+05 & 2.17 & 3.89E+04 & 2.87E+05 & 1.38 & 9.76E+04 & 1.92E+06 & 2.59 \\
1.01E+04 & 2.17E+05 & 2.10 & 3.89E+04 & 4.48E+05 & 1.67 & 9.76E+04 & 1.99E+06 & 2.63 \\
1.01E+04 & 2.13E+05 & 2.14 & 3.91E+04 & 6.07E+05 & 2.09 & 9.81E+04 & 2.35E+06 & 2.89 \\
1.01E+04 & 2.30E+05 & 2.19 & 3.92E+04 & 7.86E+05 & 2.45 & 9.91E+04 & 3.51E+06 & 3.57 \\
1.01E+04 & 2.50E+05 & 2.33 & 3.91E+04 & 9.43E+05 & 2.72 & 1.01E+05 & 4.96E+06 & 4.37 \\
1.01E+04 & 2.52E+05 & 2.32 & 3.92E+04 & 1.33E+06 & 2.90 & 1.04E+05 & 6.61E+06 & 5.03 \\
1.01E+04 & 2.96E+05 & 2.53 & 3.93E+04 & 1.33E+06 & 2.90 & 1.05E+05 & 8.08E+06 & 5.52 \\
1.01E+04 & 4.23E+05 & 3.00 & 3.97E+04 & 1.95E+06 & 3.49 &          &          &     
\end{tabular}
\end{ruledtabular}
\caption{Current study. Liquid gallium. $Pr= 0.027$, $\Gamma = 2$.}
\label{table1}
\end{table}
\end{center}

\begin{center}
\begin{table}[ht]
\begin{ruledtabular}
\begin{tabular}{ ccc |ccc |ccc }
$ \bm{Ch}$  & $\bm{Ra}$ & $\bm{Nu}$  & $ \bm{Ch}$  & $\bm{Ra}$ & $\bm{Nu}$ & $ \bm{Ch}$  & $\bm{Ra}$ & $\bm{Nu}$ \\
\hline\\[-2.5ex]
4.00E+04    & 1.5E+05 & 1.000 & 4.00E+04 & 4.0E+05 & 1.477 & 4.00E+04 & 1.50E+06 & 3.302 \\
4.00E+04    & 2.0E+05 & 1.004 & 4.01E+04 & 7.0E+05 & 2.165 & 4.00E+04 & 4.0E+06 & 5.202 \\
4.00E+04    & 3.0E+05 & 1.241 & 4.00E+04 & 1.0E+06 & 2.635 &  &  &  
\end{tabular}
\end{ruledtabular}
\caption{Current study. DNS. $Pr= 0.025$, $\Gamma = 2$.}
\label{table1b}
\end{table}
\end{center}


\begin{center}
\begin{table}[ht]
\begin{ruledtabular}
\centering
\begin{tabular}{ ccc |ccc |ccc }
$ \bm{Ch}$  & $\bm{Ra}$ & $\bm{Nu}$  & $ \bm{Ch}$  & $\bm{Ra}$ & $\bm{Nu}$ & $ \bm{Ch}$  & $\bm{Ra}$ & $\bm{Nu}$ \\
\hline\\[-2.5ex]
0        & 1.76E+06 & 6.95  & 1.11E+04 & 5.45E+07 & 16.40 & 4.43E+04 & 5.45E+07 & 16.40 \\
0        & 1.78E+06 & 6.83  & 3.83E+04 & 1.88E+06 & 4.53  & 9.13E+04 & 4.63E+06 & 4.61  \\
0        & 2.28E+06 & 7.17  & 3.84E+04 & 2.92E+06 & 5.28  & 9.13E+04 & 4.60E+06 & 4.61  \\
0        & 4.86E+06 & 8.61  & 3.86E+04 & 3.83E+06 & 6.03  & 9.17E+04 & 5.98E+06 & 5.16  \\
0        & 8.61E+06 & 9.98  & 3.86E+04 & 4.69E+06 & 6.57  & 9.21E+04 & 7.07E+06 & 5.54  \\
0        & 1.33E+07 & 11.37 & 3.89E+04 & 5.53E+06 & 7.07  & 9.15E+04 & 8.40E+06 & 5.95  \\
0        & 2.06E+07 & 13.10 & 3.85E+04 & 6.55E+06 & 7.61  & 9.10E+04 & 9.55E+06 & 6.29  \\
9.55E+03 & 1.11E+06 & 5.66  & 3.86E+04 & 7.46E+06 & 8.04  & 9.12E+04 & 1.10E+07 & 6.69  \\
9.35E+03 & 1.40E+06 & 6.10  & 3.86E+04 & 8.43E+06 & 8.48  & 9.19E+04 & 1.24E+07 & 7.06  \\
9.60E+03 & 2.84E+06 & 7.53  & 3.91E+04 & 9.74E+06 & 8.99  & 9.53E+04 & 1.86E+07 & 8.53  \\
9.65E+03 & 3.76E+06 & 8.21  & 3.86E+04 & 9.79E+06 & 8.92  & 9.44E+04 & 2.03E+07 & 8.83  \\
9.61E+03 & 4.53E+06 & 8.66  & 3.87E+04 & 1.05E+07 & 9.19  & 9.55E+04 & 2.40E+07 & 9.46  \\
9.60E+03 & 5.52E+06 & 9.04  & 3.88E+04 & 1.19E+07 & 9.59  & 9.62E+04 & 2.74E+07 & 9.99  \\
9.55E+03 & 6.39E+06 & 9.37  & 3.89E+04 & 1.26E+07 & 9.80  & 9.65E+04 & 3.05E+07 & 10.45 \\
9.58E+03 & 7.57E+06 & 9.70  & 3.88E+04 & 1.36E+07 & 10.09 & 9.82E+04 & 3.38E+07 & 10.99 \\
9.62E+03 & 8.67E+06 & 10.04 & 3.92E+04 & 1.41E+07 & 10.31 & 9.89E+04 & 3.67E+07 & 11.41 \\
9.75E+03 & 1.09E+07 & 10.52 & 3.90E+04 & 1.49E+07 & 10.40 & 1.02E+05 & 4.57E+07 & 12.60 \\
9.77E+03 & 1.15E+07 & 10.68 & 3.93E+04 & 1.64E+07 & 10.83 & 2.88E+05 & 5.98E+06 & 3.43  \\
9.94E+03 & 1.16E+07 & 10.88 & 3.95E+04 & 1.66E+07 & 10.75 & 2.75E+05 & 9.07E+06 & 4.13  \\
9.78E+03 & 1.32E+07 & 11.07 & 3.97E+04 & 1.82E+07 & 11.10 & 2.88E+05 & 1.24E+07 & 4.67  \\
9.92E+03 & 1.55E+07 & 11.49 & 3.95E+04 & 1.80E+07 & 11.22 & 2.90E+05 & 1.56E+07 & 5.38  \\
1.00E+04 & 1.88E+07 & 12.04 & 3.96E+04 & 1.94E+07 & 11.58 & 2.90E+05 & 2.07E+07 & 6.24  \\
1.01E+04 & 2.18E+07 & 12.53 & 3.98E+04 & 2.24E+07 & 12.19 & 2.90E+05 & 2.31E+07 & 6.61  \\
1.03E+04 & 2.78E+07 & 13.31 & 4.09E+04 & 2.81E+07 & 13.24 & 2.89E+05 & 2.52E+07 & 6.84  \\
1.03E+04 & 3.05E+07 & 13.68 & 4.13E+04 & 3.07E+07 & 13.69 & 2.96E+05 & 2.96E+07 & 7.34  \\
1.07E+04 & 3.91E+07 & 14.67 & 4.26E+04 & 3.90E+07 & 14.82 & 2.97E+05 & 3.34E+07 & 7.83  \\
1.09E+04 & 4.68E+07 & 15.53 & 4.31E+04 & 4.67E+07 & 15.55 & 2.88E+05 & 2.51E+07 & 6.86 
\end{tabular}
\end{ruledtabular}
\caption{Current study. Liquid gallium. $Pr= 0.027$, $\Gamma = 1$. }
\label{table2}
\end{table}
\end{center}

\begin{center}
\begin{table}[ht]
\begin{ruledtabular}
\centering
\begin{tabular}{ ccc |ccc |ccc }
$ \bm{Ch}$  & $\bm{Ra}$ & $\bm{Nu}$  & $ \bm{Ch}$  & $\bm{Ra}$ & $\bm{Nu}$ & $ \bm{Ch}$  & $\bm{Ra}$ & $\bm{Nu}$ \\
\hline\\[-2.5ex]
0 & 8.06E+06 & 8.676  & 0        & 2.53E+09 & 48.527 & 2.00E+06 & 4.99E+08 & 15.051 \\
0 & 1.27E+07 & 9.620  & 0        & 2.67E+09 & 51.986 & 2.00E+06 & 6.77E+08 & 17.400 \\
0 & 1.82E+07 & 10.667 & 7.22E+05 & 2.97E+07 & 4.103  & 2.00E+06 & 1.10E+09 & 21.055 \\
0 & 2.35E+07 & 11.930 & 7.22E+05 & 5.06E+07 & 5.545  & 2.00E+06 & 1.57E+09 & 24.800 \\
0 & 3.71E+07 & 13.228 & 7.22E+05 & 7.22E+07 & 6.936  & 2.00E+06 & 1.83E+09 & 26.111 \\
0 & 3.81E+07 & 12.562 & 7.22E+05 & 9.08E+07 & 8.382  & 2.00E+06 & 2.07E+09 & 27.000 \\
0 & 5.30E+07 & 14.542 & 7.22E+05 & 1.14E+08 & 9.703  & 3.93E+06 & 3.76E+07 & 1.000  \\
0 & 6.83E+07 & 15.986 & 7.22E+05 & 1.67E+08 & 12.033 & 3.93E+06 & 4.97E+07 & 1.578  \\
0 & 1.08E+08 & 17.879 & 7.22E+05 & 2.21E+08 & 14.050 & 3.93E+06 & 6.41E+07 & 1.940  \\
0 & 1.54E+08 & 18.989 & 7.22E+05 & 2.85E+08 & 15.578 & 3.93E+06 & 6.74E+07 & 1.875  \\
0 & 2.09E+08 & 21.055 & 7.22E+05 & 4.17E+08 & 18.989 & 3.93E+06 & 9.37E+07 & 2.930  \\
0 & 2.44E+08 & 21.055 & 7.22E+05 & 5.81E+08 & 20.696 & 3.93E+06 & 1.24E+08 & 3.998  \\
0 & 2.77E+08 & 22.556 & 7.22E+05 & 9.66E+08 & 25.445 & 3.93E+06 & 1.24E+08 & 4.000  \\
0 & 3.40E+08 & 23.347 & 7.22E+05 & 1.34E+09 & 29.455 & 3.93E+06 & 1.55E+08 & 5.001  \\
0 & 4.98E+08 & 25.011 & 2.00E+06 & 1.80E+07 & 1.120  & 3.93E+06 & 1.86E+08 & 5.940  \\
0 & 5.11E+08 & 25.665 & 2.00E+06 & 2.14E+07 & 1.448  & 3.93E+06 & 2.58E+08 & 7.691  \\
0 & 5.80E+08 & 26.564 & 2.00E+06 & 2.62E+07 & 1.970  & 3.93E+06 & 3.33E+08 & 9.290  \\
0 & 6.59E+08 & 27.259 & 2.00E+06 & 4.14E+07 & 2.958  & 3.93E+06 & 4.18E+08 & 10.667 \\
0 & 7.67E+08 & 27.972 & 2.00E+06 & 6.06E+07 & 3.960  & 3.93E+06 & 5.53E+08 & 13.800 \\
0 & 8.72E+08 & 28.458 & 2.00E+06 & 9.34E+07 & 5.358  & 3.93E+06 & 5.97E+08 & 13.228 \\
0 & 9.90E+08 & 28.952 & 2.00E+06 & 9.58E+07 & 5.180  & 3.93E+06 & 7.90E+08 & 15.400 \\
0 & 1.28E+09 & 31.284 & 2.00E+06 & 1.17E+08 & 6.419  & 3.93E+06 & 1.22E+09 & 19.319 \\
0 & 1.57E+09 & 32.103 & 2.00E+06 & 1.47E+08 & 7.490  & 3.93E+06 & 1.69E+09 & 22.600 \\
0 & 1.87E+09 & 35.904 & 2.00E+06 & 2.10E+08 & 9.375  & 3.93E+06 & 1.97E+09 & 23.147 \\
0 & 2.29E+09 & 36.843 & 2.00E+06 & 2.78E+08 & 11.000 & 3.93E+06 & 2.24E+09 & 24.600 \\
0 & 2.35E+09 & 40.155 & 2.00E+06 & 3.50E+08 & 12.455 & 3.93E+06 & 2.54E+09 & 25.887 \\
0 & 2.41E+09 & 43.764 & 2.00E+06 & 4.99E+08 & 15.600 &          &          &         
\end{tabular}
\end{ruledtabular}
\caption{\citet{cioni2000effect}. Liquid mercury. $Pr= 0.025$, $\Gamma = 1$. }
\label{table7}
\end{table}
\end{center}

\begin{center}
\begin{table}[ht]
\begin{ruledtabular}
\centering
\begin{tabular}{ ccc |ccc |ccc }
$ \bm{Ch}$  & $\bm{Ra}$ & $\bm{Nu}$  & $ \bm{Ch}$  & $\bm{Ra}$ & $\bm{Nu}$ & $ \bm{Ch}$  & $\bm{Ra}$ & $\bm{Nu}$ \\
\hline\\[-2.5ex]
0 & 7.21E+02 & 1.00 & 0 & 5.76E+03 & 1.27 & 670 & 1.52E+03 & 1.00 \\
0 & 8.18E+02 & 1.01 & 0 & 6.13E+03 & 1.34 & 670 & 2.74E+03 & 1.01 \\
0 & 1.12E+03 & 1.01 & 0 & 7.09E+03 & 1.38 & 670 & 4.24E+03 & 1.00 \\
0 & 1.30E+03 & 1.01 & 0 & 6.57E+03 & 1.40 & 670 & 6.00E+03 & 1.01 \\
0 & 1.46E+03 & 1.02 & 0 & 8.49E+03 & 1.44 & 670 & 7.04E+03 & 1.01 \\
0 & 1.86E+03 & 1.03 & 0 & 9.10E+03 & 1.48 & 670 & 9.17E+03 & 1.01 \\
0 & 2.12E+03 & 1.03 & 0 & 1.05E+04 & 1.49 & 670 & 8.26E+03 & 1.01 \\
0 & 2.27E+03 & 1.04 & 0 & 1.07E+04 & 1.54 & 670 & 1.16E+04 & 1.03 \\
0 & 2.50E+03 & 1.05 & 0 & 1.11E+04 & 1.55 & 670 & 1.03E+04 & 1.02 \\
0 & 3.35E+03 & 1.11 & 0 & 1.31E+04 & 1.61 & 670 & 1.25E+04 & 1.03 \\
0 & 3.52E+03 & 1.14 & 0 & 1.12E+04 & 1.63 & 670 & 1.34E+04 & 1.04 \\
0 & 3.77E+03 & 1.16 & 0 & 1.20E+04 & 1.64 & 670 & 1.37E+04 & 1.05 \\
0 & 4.15E+03 & 1.20 & 0 & 1.30E+04 & 1.68 & 670 & 1.44E+04 & 1.05 \\
0 & 4.71E+03 & 1.23 & 0 & 1.32E+04 & 1.71 & 670 & 1.48E+04 & 1.06 \\
0 & 4.55E+03 & 1.25 & 0 & 1.48E+04 & 1.71 & 670 & 1.53E+04 & 1.08 \\
0 & 5.22E+03 & 1.25 &   &          &      &     &          &     
\end{tabular}
\end{ruledtabular}
\caption{\citet{aurnou2001experiments}. Liquid gallium. $Pr= 0.023$, $\Gamma = 6$. }
\label{table5}
\end{table}
\end{center}

\begin{center}
\begin{table}[ht]
\begin{ruledtabular}
\centering
\begin{tabular}{ ccc |ccc |ccc }
$ \bm{Ch}$  & $\bm{Ra}$ & $\bm{Nu}$  & $ \bm{Ch}$  & $\bm{Ra}$ & $\bm{Nu}$ & $ \bm{Ch}$  & $\bm{Ra}$ & $\bm{Nu}$ \\
\hline\\[-2.5ex]
0 & 1.74E+03 & 1.05 & 0   & 2.65E+04 & 1.80 & 400  & 4.92E+04 & 1.48 \\
0 & 2.29E+03 & 1.06 & 0   & 3.64E+04 & 1.95 & 400  & 6.13E+04 & 1.60 \\
0 & 3.08E+03 & 1.10 & 0   & 4.59E+04 & 2.08 & 400  & 7.39E+04 & 1.69 \\
0 & 4.72E+03 & 1.23 & 0   & 5.58E+04 & 2.19 & 400  & 8.46E+04 & 1.78 \\
0 & 6.92E+03 & 1.35 & 400 & 9.27E+03 & 1.00 & 1600 & 3.22E+04 & 1.01 \\
0 & 9.19E+03 & 1.39 & 400 & 1.49E+04 & 1.05 & 1600 & 4.56E+04 & 1.07 \\
0 & 1.07E+04 & 1.45 & 400 & 2.11E+04 & 1.12 & 1600 & 6.22E+04 & 1.18 \\
0 & 1.49E+04 & 1.59 & 400 & 2.68E+04 & 1.20 & 1600 & 7.60E+04 & 1.29 \\
0 & 1.92E+04 & 1.67 & 400 & 3.65E+04 & 1.30 & 1600 & 9.10E+04 & 1.40
\end{tabular}
\end{ruledtabular}
\caption{\citet{burr2001rayleigh}. Liquid Na-K alloy. $0.017<Pr<0.021$, rectangular box $20:10:1$. }
\label{table6}
\end{table}
\end{center}

\begin{center}
\begin{table}[ht]
\begin{ruledtabular}
\centering
\begin{tabular}{ ccc |ccc |ccc }
$ \bm{Ch}$  & $\bm{Ra}$ & $\bm{Nu}$  & $ \bm{Ch}$  & $\bm{Ra}$ & $\bm{Nu}$ & $ \bm{Ch}$  & $\bm{Ra}$ & $\bm{Nu}$ \\
\hline\\[-2.5ex]
9.46E+03 & 2.37E+06 & 7.23  & 4.74E+04 & 8.44E+06 & 8.40  & 2.85E+05 & 1.09E+07 & 4.89  \\
9.50E+03 & 3.30E+06 & 7.78  & 9.35E+04 & 3.81E+06 & 4.50  & 2.85E+05 & 1.34E+07 & 5.29  \\
9.54E+03 & 4.32E+06 & 8.31  & 9.39E+04 & 5.02E+06 & 5.11  & 9.35E+05 & 7.39E+06 & 2.31  \\
9.60E+03 & 5.95E+06 & 8.94  & 9.43E+04 & 6.33E+06 & 5.67  & 9.40E+05 & 9.80E+06 & 2.61  \\
9.58E+03 & 7.53E+06 & 9.42  & 9.49E+04 & 8.25E+06 & 6.46  & 9.46E+05 & 1.23E+07 & 2.91  \\
9.58E+03 & 1.06E+07 & 10.20 & 9.48E+04 & 1.01E+07 & 7.05  & 9.54E+05 & 1.59E+07 & 3.37  \\
9.74E+03 & 2.80E+07 & 13.00 & 9.47E+04 & 1.33E+07 & 8.11  & 9.54E+05 & 1.90E+07 & 3.76  \\
1.01E+04 & 5.08E+07 & 15.00 & 9.53E+04 & 1.89E+07 & 9.56  & 9.57E+05 & 2.53E+07 & 4.29  \\
4.68E+04 & 3.10E+06 & 5.52  & 9.96E+04 & 5.40E+07 & 14.10 & 9.67E+05 & 3.48E+07 & 5.25  \\
4.70E+04 & 4.11E+06 & 6.24  & 2.80E+05 & 5.23E+06 & 3.27  & 1.02E+06 & 8.85E+07 & 8.84  \\
4.72E+04 & 5.13E+06 & 6.99  & 2.81E+05 & 6.70E+06 & 3.83  & 1.24E+06 & 1.70E+08 & 11.90 \\
4.75E+04 & 6.81E+06 & 7.81  & 2.83E+05 & 8.37E+06 & 4.28  &          &          &      
\end{tabular}
\end{ruledtabular}
\caption{\citet{king2015magnetostrophic}. Liquid gallium. $Pr= 0.025$, $\Gamma = 1$. }
\label{table4}
\end{table}
\end{center}

\begin{center}
\begin{table}[ht]
\begin{ruledtabular}
\centering
\begin{tabular}{ ccc |ccc |ccc }
$ \bm{Ch}$  & $\bm{Ra}$ & $\bm{Nu}$  & $ \bm{Ch}$  & $\bm{Ra}$ & $\bm{Nu}$ & $ \bm{Ch}$  & $\bm{Ra}$ & $\bm{Nu}$ \\
\hline\\[-2.5ex]
1.73E+02 & 4.18E+06 & 7.59  & 1.73E+04 & 2.51E+07 & 11.80 & 2.12E+05 & 3.34E+07 & 8.84  \\
1.73E+02 & 6.27E+06 & 8.10  & 1.73E+04 & 3.34E+07 & 12.30 & 2.12E+05 & 5.85E+07 & 12.00 \\
1.73E+02 & 1.05E+07 & 9.19  & 1.74E+04 & 5.78E+07 & 15.50 & 4.32E+05 & 1.05E+07 & 3.18  \\
1.73E+02 & 1.67E+07 & 11.00 & 1.74E+04 & 5.82E+07 & 15.50 & 4.32E+05 & 1.05E+07 & 3.67  \\
1.72E+02 & 3.34E+07 & 12.50 & 1.73E+04 & 5.84E+07 & 15.00 & 4.32E+05 & 1.05E+07 & 3.18  \\
1.74E+02 & 5.82E+07 & 15.70 & 3.89E+04 & 1.05E+07 & 7.67  & 4.32E+05 & 1.05E+07 & 3.41  \\
1.78E+02 & 5.91E+07 & 15.70 & 6.92E+04 & 6.26E+06 & 5.48  & 4.32E+05 & 1.67E+07 & 4.98  \\
4.32E+03 & 4.18E+06 & 7.49  & 6.92E+04 & 1.05E+07 & 6.62  & 4.33E+05 & 3.34E+07 & 7.20  \\
4.32E+03 & 6.27E+06 & 7.71  & 6.92E+04 & 1.67E+07 & 8.55  & 4.33E+05 & 5.86E+07 & 10.10 \\
4.32E+03 & 1.05E+07 & 9.23  & 6.92E+04 & 2.51E+07 & 9.77  & 7.31E+05 & 1.67E+07 & 3.63  \\
4.33E+03 & 1.67E+07 & 10.90 & 6.93E+04 & 3.34E+07 & 11.00 & 7.31E+05 & 3.34E+07 & 5.90  \\
4.33E+03 & 2.51E+07 & 11.60 & 6.93E+04 & 3.34E+07 & 10.80 & 7.32E+05 & 5.86E+07 & 8.89  \\
4.33E+03 & 3.34E+07 & 12.30 & 6.94E+04 & 5.84E+07 & 14.40 & 1.11E+06 & 1.67E+07 & 2.97  \\
4.34E+03 & 5.81E+07 & 15.20 & 6.94E+04 & 5.85E+07 & 14.00 & 1.11E+06 & 3.34E+07 & 4.75  \\
1.73E+04 & 4.18E+06 & 5.99  & 6.94E+04 & 5.85E+07 & 14.40 & 1.11E+06 & 3.34E+07 & 5.12  \\
1.73E+04 & 6.27E+06 & 7.28  & 6.94E+04 & 5.86E+07 & 14.30 & 1.11E+06 & 3.34E+07 & 4.88  \\
1.73E+04 & 1.05E+07 & 9.07  & 2.12E+05 & 1.05E+07 & 5.05  & 1.11E+06 & 5.85E+07 & 7.63  \\
1.73E+04 & 1.67E+07 & 10.90 & 2.12E+05 & 1.67E+07 & 6.75  &          &          &      
\end{tabular}
\end{ruledtabular}
\caption{\citet{zurner2020flow}. Liquid GaInSn. $Pr= 0.029$, $\Gamma = 1$. }
\label{table3}
\end{table}
\end{center}